\pdfoutput=1
\documentclass[a4paper,12pt]{article}

\usepackage[a4paper,text={16.8cm,22.4cm}]{geometry}
\usepackage{amsmath,amssymb,bm,psfrag,cite,graphicx}
\usepackage{mathtools}
\usepackage[normalem]{ulem}
\usepackage{slashed}

\usepackage{caption}
\usepackage{subcaption}

\usepackage{xcolor}

\usepackage[Q=yes,pverb-linebreak=no]{examplep}

\allowdisplaybreaks 
\begin{document}

\begin{titlepage}

\vspace*{1.8cm}
\begin{center}
\Large\bf
Event-Based Transverse Momentum Resummation 
\end{center}

%\vspace{0.2cm}
%\begin{center}
%Thomas Becher$^a$, Monika Hager$^a$ and name$^b$\\
%\vspace{0.4cm}
%{\sl 
%${}^a$\,Institut f\"ur Theoretische Physik, Universit\"at Bern\\
%Sidlerstrasse 5, CH--3012 Bern, Switzerland\\[0.3cm]
%%
%${}^b$\,name of institute, 
%name of university\\ 
%adress}
%\end{center}

\vspace{0.2cm}
\begin{center}
Thomas Becher and Monika Hager\\
\vspace{0.4cm}
{\sl 
Albert Einstein Center for Fundamental Physics, Institut f\"ur Theoretische Physik,\\ 
Universit\"at Bern, Sidlerstrasse 5, 3012 Bern, Switzerland}
\end{center}

\vspace{0.2cm}
\begin{abstract}
\label{sec:abstract}
\vspace{0.2cm}
\noindent 

We have developed a framework for automated transverse momentum resummation for arbitrary electroweak 
final states based on reweighting tree-level events. 
It is fully differential in the kinematics of the electroweak final states, 
which facilitates a straightforward analysis of arbitrary observables 
in the small transverse momentum region. 
We have implemented the resummation at next-to-next-to-leading logarithmic accuracy 
and match to next-to-leading fixed-order results using the event generator {\sc MadGraph5\Q{_}aMC@NLO}. 
Results for $Z$ and $W$ boson production with leptonic decay as well as $WZ$ production are presented. 
We compare to experimental measurements for the transverse momentum and the angular observable $\phi^*$. 

\end{abstract}
\vfil

\end{titlepage}

\section{Introduction}
\label{sec:intro}

Since the emission of particles with large transverse momentum $q_T$ is suppressed by the strong coupling $\alpha_s(q_T)$, most of the cross section at hadron colliders arises from events  with low-$q_T$ radiation. This is true in particular for the production of electroweak bosons, i.e.\ $Z$'s, $W^\pm$'s and Higgs bosons. With its large data sets, the LHC can measure transverse momentum spectra of such bosons with exquisite precision and these results are used, for example, to determine the $W$ mass, as first achieved at the LHC in \cite{Aaboud:2017svj}. For this determination, the region of small $q_T$ is especially important.

The suppression of radiation by the coupling constant becomes ineffective at low transverse momentum, since it gets compensated by large logarithms of the ratio of the transverse momentum to the invariant mass of the electroweak boson. The all-order structure of these enhanced contributions was first understood  by Collins, Soper and Sterman (CSS) \cite{Collins:1984kg}, who showed that the cross section factorizes in (transverse) position space into a product of a hard function which encodes the virtual contributions to the electroweak boson production process and collinear functions describing the QCD emissions at low transverse momentum. The hard function depends on the electroweak process under consideration, while the collinear functions are universal and only distinguish quark-induced from gluon-induced processes. The result of CSS has been implemented by different authors and has also been rederived in the context of Soft-Collinear Effective Theory (SCET) \cite{Bauer:2000yr,Bauer:2001yt,Beneke:2002ph} (see \cite{Becher:2014oda,Becher:2018gno,Cohen:2019wxr} for reviews). The result in SCET \cite{Becher:2010tm,Chiu:2012ir} makes it clear that the process involves two distinct sources of large logarithms: i.) logarithms due to the different scales associated with the hard process and the radiation and ii.) logarithms due to the rapidity difference in the low-$q_T$ emissions from the partons flying along the beams to the left and right  (these rapidity logarithms were not addressed in earlier work on transverse momentum resummation within SCET \cite{Gao:2005iu,Idilbi:2005er,Mantry:2009qz}). The first kind of logarithms are resummed by standard renormalization-group (RG) methods, while the second class is either exponentiated directly, using the collinear anomaly formalism \cite{Becher:2010tm}, or resummed via a dedicated rapidity RG \cite{Chiu:2011qc,Chiu:2012ir}. While the factorization holds in position space,  recently also methods to resum directly in momentum space have been developed \cite{Ebert:2016gcn, Monni:2016ktx}. A number of computer codes for the resummation are available, e.g.\ {\sc RESBOS}\cite{Balazs:1997xd}, {\sc CuTe} \cite{Becher:2012yn}, {\sc DYRes} \cite{Catani:2015vma}, {\sc MATRIX} \cite{Grazzini:2015wpa,Grazzini:2017mhc} and {\sc RadISH} \cite{Bizon:2017rah}. For single-boson processes, the resummation is now performed up to next-to-next-to-next-to-logarithmic (NNNLL) accuracy \cite{Becher:2012yn,Bizon:2017rah,Bizon:2018foh,Chen:2018pzu}.

In the present paper, we present an efficient and flexible framework which achieves the following three goals:
\begin{enumerate}
\item It performs the resummation for arbitrary electroweak final states.
\item It computes any hadronically inclusive observable dominated by low-$q_T$ radiation and can take into account experimental cuts on the final-state leptons.
\item It automatically matches the resummed predictions to fixed-order results in kinematic regions were $q_T$ becomes large.
\end{enumerate}
In order to achieve this flexibility, we make use of {\sc MadGraph5\Q{_}aMC@NLO} \cite{Alwall:2014hca}, to compute the process-specific parts of the resummed cross section and supply it with the universal ingredients needed to achieve the resummation. Since we work at low $q_T$, we are close to Born level kinematics and can use the tree-level event generator to produce the leptonic final state. The automated one-loop code included in {\sc MadGraph5\Q{_}aMC@NLO} is used to compute the virtual corrections to the hard scattering process. These results are then combined with the resummation factors and the universal collinear functions, which are tabulated and interpolated using PDF codes. More specifically, we start with tree level events which we reweight and boost to obtain resummed events. By analyzing these resummed events, we are able to impose cuts and extract arbitrary leptonic distributions. To perform the matching, we rely on the NLO fixed-order implementation of {\sc MadGraph5\Q{_}aMC@NLO}.

The approach of reweighting fixed-order results to perform resummation was pioneered in \cite{Banfi:2012du}, using MCFM \cite{Campbell:1999ah}. Our implementation of the reweighting follows \cite{Becher:2014aya}, in which we used {\sc MadGraph5\Q{_}aMC@NLO} for automated jet-veto cross section resummation. Compared to this earlier work, transverse-momentum resummation involves a number of complications, which include the Fourier inversion back to momentum space and the necessity to account for recoil effects. The framework for transverse momentum resummation we use was developed in \cite{Becher:2011xn,Becher:2012yn}. This was implemented earlier in the {\sc CuTe} code \cite{Becher:2012yn}, which was however restricted to the inclusive $q_T$ spectrum of single bosons. Our new event-based framework extends it in the ways enumerated above and also  introduces a novel efficient method to perform the matching and switch off the resummation at larger $q_T$. Our current implementation computes quark induced processes, has NNLL accuracy and matches at $\mathcal{O}(\alpha_s)$ to fixed order. Extending it to higher accuracy requires two-loop ingredients which are not universally known, but could be implemented by hand for single-boson processes and for those diboson processes where they are available. 

An important example of a kinematic quantity which correlates with
the dilepton transverse momentum is the variable $\phi^*$ introduced in \cite{Banfi:2010cf, Banfi:2012du, Sirunyan:2017igm}. 
It is defined using the directions of the final-state leptons from the decay $Z \to \ell^+ \ell^-$ as follows
\begin{equation}\label{phiDef}
\phi^* \coloneqq \tan\left( \frac{\pi - \Delta\phi}{2} \right) \sin(\theta^*)\,, \qquad \text{with}
\quad \cos(\theta^*) \coloneqq \tanh\left( \frac{\Delta\eta}{2} \right) \,.
\end{equation}
Here $\Delta\phi$ is the opening angle of the leptons in the azimuthal plane and
$\Delta\eta = \eta^- - \eta^+$ the difference in their pseudorapidity. Since only angular measurements are needed to determine $\phi^*$, this quantity can be obtained even more precisely than $q_T$, which also requires lepton-energy measurements. Once $q_T$ approaches zero, the two leptons align back-to-back in the azimuthal plane and $\phi^*$ approaches zero. Indeed, computing the double differential cross section in $q_T$ and $\phi^*$, we observe a strong correlation among the two variables. We will compare to the  $\phi^*$ measurements of ATLAS \cite{Aad:2015auj} in our paper, to illustrate our method in practice.

Our paper is organized as follows. In Section~\ref{sec:methods} we will review the factorization formula and the ingredients needed for NNLL accuracy. The implementation is discussed in Section~\ref{sec:autom}, which provides details on the treatment of recoil effects, event generation,  the matching to fixed order and the structure of the codes used to perform the resummation. In Section~\ref{sec:result}, we give numerical results for different processes, validate our results against {\sc CuTe} and compare to experimental data. Conclusions and an outlook are presented in Section~\ref{sec:concl}.

\section{Factorization at low transverse momentum}
\label{sec:methods}

We consider the scattering of protons with momenta $p_1$ and $p_2$ producing any number of massive electroweak bosons ($W^\pm$, $Z$, $H$) with momenta $q_i$, possibly decaying to leptons or photons, accompanied by hadronic radiation with total momentum $p_X$. The center-of-mass energy of the collision is $s=(p_1+p_2)^2$  and the total electroweak momentum is
\begin{equation}
q= q_1+ q_2+ \dots + q_N\, .
\end{equation}
We will analyze the cross section in the region where the transverse momentum $q_\perp$ is much smaller than the invariant mass $Q^2=q^2$ of the electroweak final state, and we use the notation $q_T = \sqrt{- q_\perp^2}$ to denote the positive scalar quantity associated with it. If we neglect the small transverse momentum, we can write the electroweak momentum as
\begin{equation}
q^\mu = \xi_1 p_1+\xi_2 p_2 + \mathcal{O}(q_\perp)\,,
\end{equation}
where the momenta $\hat p_1=\xi_1 p_1$ and $\hat p_2=\xi_2 p_2$ correspond to the large light-cone momentum components along the beam directions. 
Expanding the cross section around $q_T=0$ one obtains the factorization formula \cite{Becher:2010tm,Becher:2011xn,Becher:2012yn}
\begin{equation}\label{eq:sigma}
\begin{aligned}
  d\sigma =& \sum_{ij \in\{q,\bar{q},g\}} \int_0^1 \!d\xi_1 \int_0^1\! d\xi_2\,  d\sigma^0_{ij}( \hat{p}_1, \hat{p}_2, q_1, ... , q_N) \, {\cal H}_{ij}( \hat{p}_1, \hat{p}_2, q_1, ... , q_N, \mu) \\
  &\;\;\;\times \frac1{4\pi} \int_{-\infty}^{\infty} d^2x_{\perp}\,
  e^{-iq_{\perp}\cdot x_{\perp}} 
  \left(\frac{x_T^2Q^2}{b_0^2}\right)^{-F_{ij}(x_\perp,\mu)}
  { B}_i(\xi_1, x_{\perp},\mu)\,{ B}_j(\xi_2, x_{\perp},\mu) \, ,\,
\end{aligned}
\end{equation}
which is equivalent to the CSS result  \cite{Collins:1984kg}. We sum over partonic channels and integrate over the momentum fractions $\xi_1$ and $\xi_2$ of the partons entering the hard scattering process. We have introduced the abbreviation $b_0 = 2e^{-\gamma_E}$, where $\gamma_E$ is the Euler-Mascheroni constant. The formula, whose ingredients will be discussed below, involves a Fourier convolution over the transverse separation $x_\perp$ and holds up to terms  suppressed by powers of $q_T^2/Q^2$. The cross section $d\sigma$ is inclusive in the hadronic radiation but completely differential in the electroweak momenta $q_1, \dots, q_N$. To compute a specific cross section, such as the transverse momentum spectrum, one imposes suitable constraints on these momenta and integrates \eqref{eq:sigma} over the electroweak phase space. 
%For brevity we suppress in the following the momentum dependence of the Born level partonic cross section $d\sigma^0_{ij}\equiv d\sigma^0_{ij}(\hat p_1, \hat p_2, q_1, ... , q_N)$. 

\begin{figure}[t!]
\centering
\includegraphics[width=0.85\textwidth]{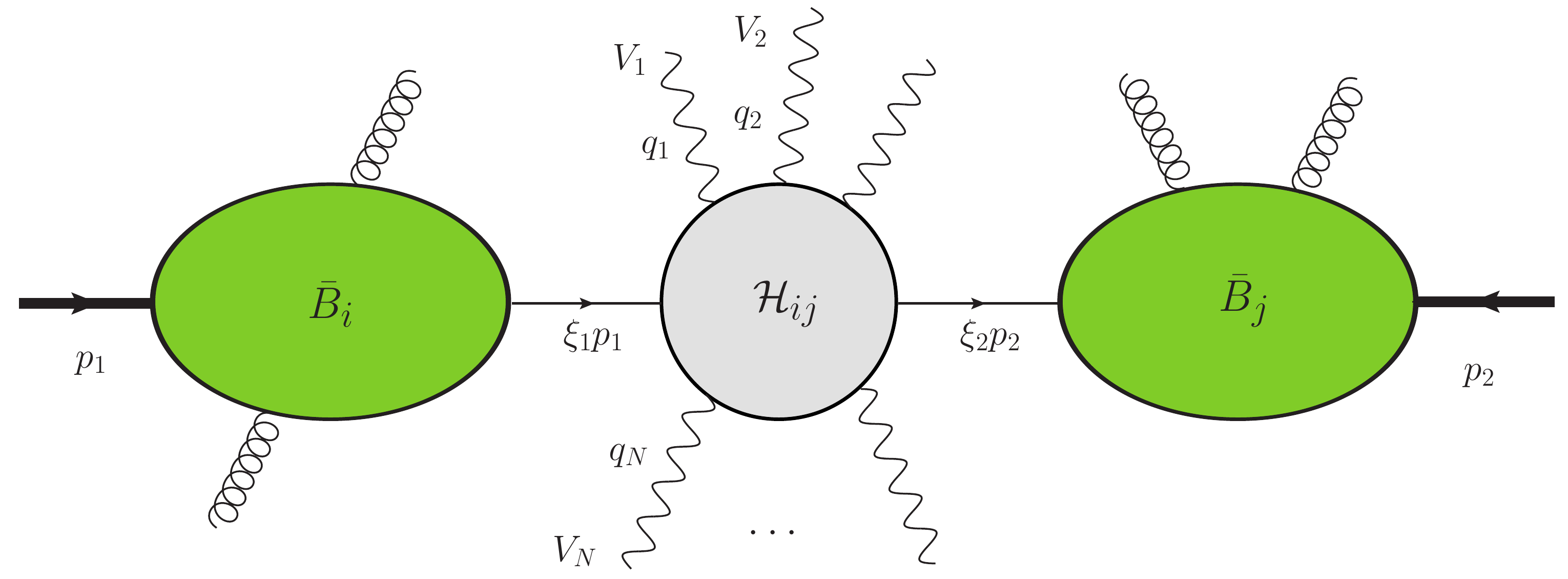}
\caption{Structure and kinematics of the factorization theorem for electroweak boson production at low transverse momentum. The wavy lines denote the bosons in the final state. We can also include their leptonic decays in our framework.}
\label{fig:WW1}
\end{figure}

\begin{figure}[t!]
\centering
\includegraphics[width=0.5\textwidth]{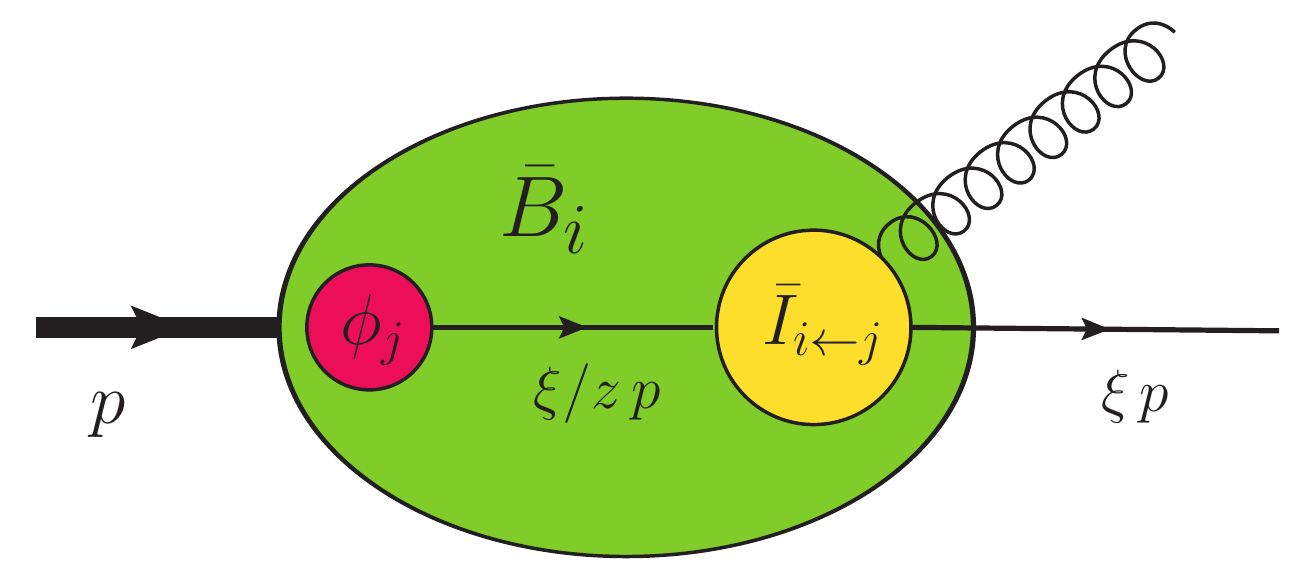}
\caption{Schematic representation of the beam functions that encode the collinear emissions.}
\label{fig:WW2}
\end{figure}

The structure of the cross section  \eqref{eq:sigma} is shown graphically in Figure~\ref{fig:WW1} and is similar to the leading order cross section which reads
\begin{equation}\label{eq:sigma0}
\begin{aligned}
  d\sigma^{\rm LO}
  &= \sum_{ij \in\{\bar{q},\bar{q},g\}} \int_0^1 \!d\xi_1 \int_0^1\! d\xi_2\, d\sigma^0_{ij}(\hat p_1, \hat p_2, q_1, ... , q_N)\,\phi_i(\xi_1,\mu)\,\phi_j(\xi_2, \mu) \,.
\end{aligned}
\end{equation} 
Compared to the Born level result, the cross section \eqref{eq:sigma} involves two additional ingredients. First of all, the resummed result involves a Fourier convolution with two beam functions ${ B}_i(\xi, x_{\perp},\mu)$ instead of a convolution with ordinary parton distribution functions (PDFs) $\phi_i(\xi,\mu)$. The beam functions describe the soft and collinear QCD emissions which accompany the incoming parton, see Figures~ \ref{fig:WW1} and \ref{fig:WW2}. We will discuss these functions and the associated Fourier integral over the transverse separation $x_\perp$ in more detail below. Let us note that for gluon-induced processes, such as Higgs production, two beam function structures arise. In this case the factorization formula involves a sum of two products of beam functions rather than just a product \cite{Catani:2010pd,Becher:2012yn}. However, the second structure first arises at NNNLL and is thus not relevant in the present paper.

Secondly, the resummed result also includes the virtual corrections to the Born level process. These are part of the hard function ${\cal H}_{ij}$, which is given by the loop contribution to the process, after subtracting its divergences in $\overline{\rm MS}$ renormalization. We write the expansion of the hard function in the form
\begin{equation}
\mathcal{H}_{ij}(\hat{p}_1, \hat p_2, q_1, ... , q_N,\mu) 
    = 1+ \frac{\alpha_s(\mu)}{4\pi}  \mathcal{H}^{(1)}_{ij}(\hat{p}_1, \hat p_2, q_1, ... , q_N,\mu)  + \mathcal{O}(\alpha_s^2)\,.
\end{equation} 
 The one-loop hard function for quark-induced processes takes the form
\begin{equation}\label{eq:oneloopH}
    \mathcal{H}^{(1)}_{q\bar{q}} = -2 C_F  \ln^2\frac{Q^2}{\mu^2}+ 6 C_F \ln\frac{Q^2}{\mu^2} + h_0(\hat p_1, \hat p_2, q_1, ... , q_N) \,.
\end{equation}
The $\mu$ dependence is universal since it is driven by the anomalous dimension of the operator with a single collinear quark field for each beam direction. All nontrivial information about the process resides in the scale independent piece $h_0$. For $Z$ boson production we have $h_0=C_F(-16 + 7 \pi^2/3)$. For more complicated processes, we use {\sc MadGraph5\Q{_}aMC@NLO} to compute the one-loop corrections, as described in detail in \cite{Becher:2014aya}. Specifically, running the code at an arbitrary reference scale $\mu_{\rm Mad}$, the hard function is related to the finite part $C_0$ of the virtual contribution obtained from {\sc MadGraph5\Q{_}aMC@NLO} as follows:  
\begin{equation}
   h_0(\hat p_1, \hat p_2, q_1, ... , q_N)
   = 2 C_0(\hat p_1, \hat p_2, q_1, ... , q_N,\mu_{\rm Mad}) + C_F \left[ \frac{\pi^2}{3} +  2\ln^2\frac{Q^2}{\mu_{\rm Mad}^2} -6  \ln\frac{Q^2}{\mu_{\rm Mad}^2} \right]\, .
\end{equation}
We observe that \eqref{eq:oneloopH} suffers from large logarithms when $\mu^2\ll Q^2$, while the beam functions will involve large logarithms for  $\mu^2 \gg q_T^2$. To avoid this problem, we solve the RG equation of the hard function to evolve it to low values of $\mu$ at which the beam function is free of large logarithms. The result then takes the form 
\begin{equation}\label{eq:resHard}
\mathcal{H}_{q\bar q}(\hat{p}_1, \hat p_2, q_1, ... , q_N,\mu) = U(Q^2, \mu_h,\mu) \,\mathcal{H}_{q\bar q}(\hat{p}_1, \hat p_2, q_1, ... , q_N,\mu_h)  \,,
\end{equation}
and we choose the starting scale of the evolution to be $\mu_h \sim Q$. The analytical expression for the evolution factor $U(Q^2, \mu_h,\mu)$ is given in Appendix \ref{app:c}. 

Let us now discuss the Fourier integral. Despite the fact that it describes low-energy dynamics, the integral depends on the large scale $Q^2$ through the collinear anomaly \cite{Becher:2010tm}. This dependence exponentiates in \eqref{eq:sigma} and is driven by the anomaly exponent $F_{ij}$, that was derived to two loops in \cite{Becher:2010tm} and has now even been determined at $\mathcal{O}(\alpha_s^3)$ in \cite{Li:2016ctv,Vladimirov:2016dll}. The beam functions ${ B}_i$ are given by a convolution of a perturbative part, describing collinear and soft emissions at small transverse momentum, with the usual PDFs. The beam functions are illustrated in Figure~\ref{fig:WW2} and will be discussed in detail below.

In perturbation theory, the functions  ${ B}_i$ are polynomials in the logarithm
\begin{equation}
L_\perp = \ln\frac{x_T^2\mu^2}{b_0^2} \,,
\end{equation}
and it is useful to follow  \cite{Becher:2011xn} and factor out their double logarithmic dependence by rewriting
\begin{equation}\label{exph}
{ B}_i(\xi_i,x_\perp,\mu)
= e^{h_i(L_\perp,a_s)}\,{\bar{B}}_i(\xi_i,x_\perp,\mu) \,,
\end{equation}
where we have introduced the abbreviation $a_s  = \alpha_s(\mu)/4\pi$. The double-logarithmic exponent $h_i(L_\perp,a_s)$ is defined as the solution of the RG equation
\begin{equation}
\frac{d}{d\ln\mu} h_i(L_\perp,a_s) =  C_i \,\gamma_{\rm cusp} \, L_\perp - 2 \gamma^i(a_s)
\end{equation}
with boundary condition $h_i(0,a_s)=0$.
% so that
%\begin{equation}
%h_i(L_\perp,a_s) = a_s \left[ C_i \Gamma_0\, L_\perp^2 - \gamma_0^i L_\perp \right]
%\end{equation} 
%at one loop. 
For quark-induced processes, we have $C_i=C_F$, while $C_i=C_A$ in the gluon case. The functions ${\bar{B}}_i$ are single logarithmic and it is convenient to combine the double logarithmic part with the anomaly into a single exponent
\begin{equation}\label{combi}
e^{g_i(\eta_i,L_\perp,a_s)}
= \left(\frac{x_T^2Q^2}{b_0^2}\right)^{-F_{ij}(L_{\perp},a_s)} e^{h_i(L_\perp,a_s)} e^{h_j(L_\perp,a_s)}\,.
\end{equation}
While the beam functions are flavor dependent because they contain the PDFs, the exponents in this equation only depend on the color representation of the partons entering the hard scattering, i.e. they only differ between the quark case (fundamental representation, $g_i = g_F$) and the gluon induced process (adjoint representation, $g_i = g_A$). We list the exponent $g_i$ in the appendix  in (\ref{logE}), it was first given in \cite{Becher:2011xn}. The exponent depends on  the variable
\begin{equation}\label{etadef}
   \eta_i\equiv \eta_i(Q^2,\mu) = \frac{C_i\alpha_s(\mu)}{\pi}\,\ln\frac{Q^2}{\mu^2} \sim 1\,,
\end{equation}
which captures the large anomaly logarithms. 

The Fourier integral in the factorization formula has some remarkable properties at very low transverse momentum. One would naively expect that the relevant scale for the integral tends to zero as $q_T \to 0$ but this is not the case, as was noted by Parisi and Petronzio \cite{Parisi:1979se} already before the all-order factorization of the cross section was fully understood. For very low $q_T$, the Fourier factor becomes ineffective and it is instead the Sudakov double logarithms inside the exponent $g_i$ which regularize the integration of the transverse separation. Analyzing the corresponding Gaussian integral, one finds that the associated scale $q_*$ is given by the value of $\mu$ at which $\eta_i$ becomes equal to one \cite{Becher:2011xn}
\begin{equation}\label{qstar}
   q_*^2= Q^2\,\exp\left( - \frac{\pi}{C_i \,\alpha_s(q_*)} \right). 
\end{equation}
For $Z$ production $q_* \approx 1.88\,{\rm GeV}$. In our numerical work, we therefore use $\mu=q_T+q_*$ as the default choice for the factorization scale. A consequence of the appearance of the dynamical scale $q_*$ is that the logarithm $L_\perp$, which usually counts as an $\mathcal{O}(1)$ quantity, must be counted as $L_\perp\sim \frac{1}{\sqrt{\alpha_s}}$ for $q_T \to 0$ \cite{Becher:2011xn}. One must therefore resum terms of the form $\alpha_s^n L_\perp^{2n}$, which now count as $\mathcal{O}(1)$. In \eqref{exph}, we have achieved this resummation by pulling out the factor $h_i$ from the beam functions and exponentiating it. The exponent $g_i(\eta_i,L_\perp,a_s)$ given in  (\ref{logE}) contains all necessary terms to achieve $\mathcal{O}(\alpha_s)$ accuracy also in the counting relevant for $q_T \to 0$. The appearance of the dynamical scale $q_*$ can also be understood from a momentum space perspective. Instead of soft radiation recoiling against the weak boson, the typical radiation for $q_T \to 0$ consists of QCD emissions at a scale $q_*$ recoiling against each other. This is the physical picture which underlies the momentum space formalism proposed in \cite{Monni:2016ktx}.

\begin{figure}
\centering
\begin{tabular}{cc}
 \includegraphics[height=0.14\textwidth]{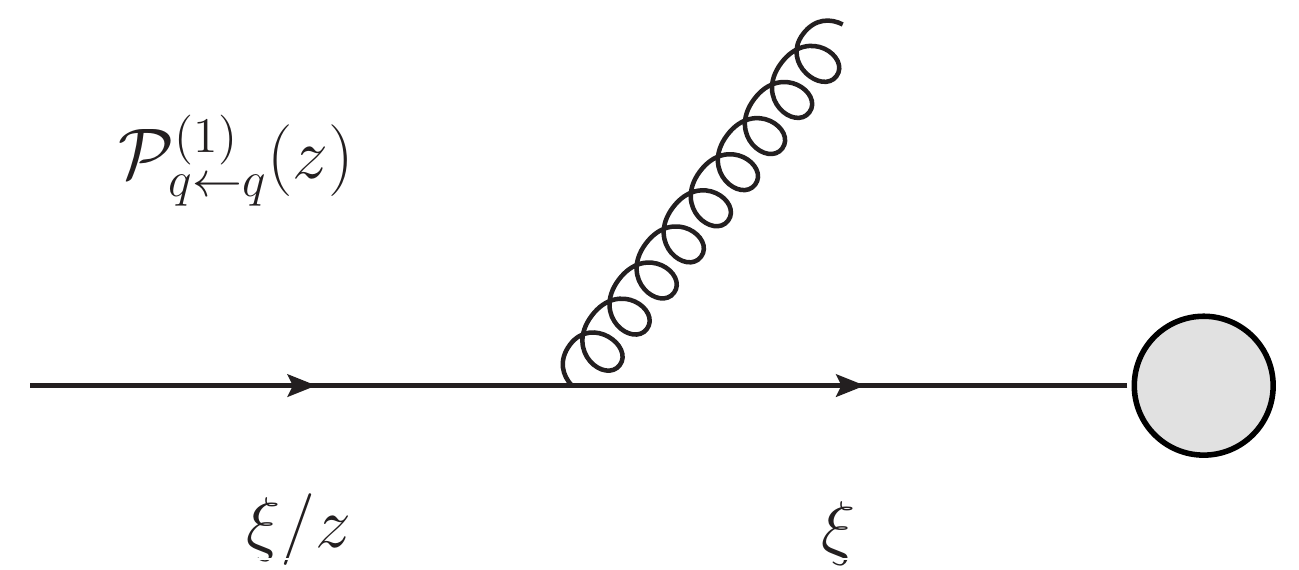}  &  \includegraphics[height=0.14\textwidth]{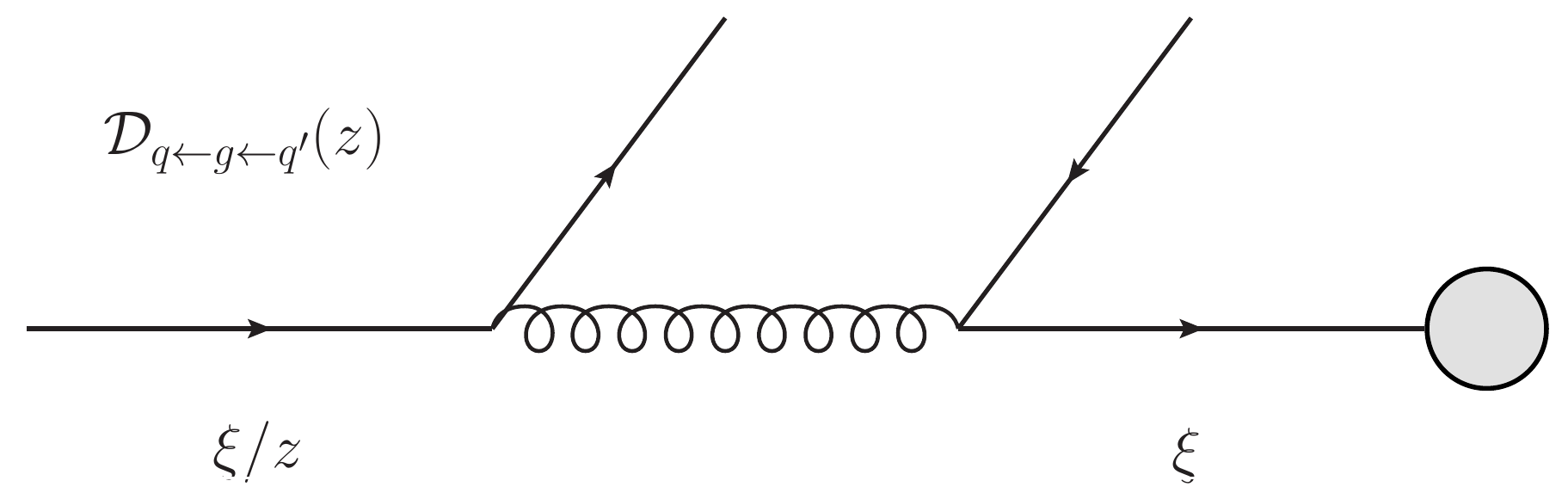}
 \end{tabular}
\caption{Sample one- and two-loop ingredients to the beam functions. The incoming parton is on the left and the grey blob indicates the hard interaction.
\label{fig:PandDterms}}
\end{figure}

As depicted in Figure~\ref{fig:WW2}, the transverse-position dependent beam function $ {\bar{B}}_i$ factorizes into a perturbative kernel $\bar{I}_{i\leftarrow j}$ describing the soft and collinear emissions at low transverse momentum, with the PDFs
\begin{equation}\label{BfunI}
    {\bar{B}}_i(\xi,x_\perp,\mu) 
   = \sum_j \int_\xi^1 \frac{dz}{z}\,\bar{I}_{i\leftarrow j}(z,x_\perp,\mu)\,
    \phi_j(\xi/z,\mu) \,, 
\end{equation}
For NNLL resummation, we need the one-loop result for $\bar{I}_{i\leftarrow j}$  which takes the form \cite{Becher:2010tm,Becher:2011xn}
\begin{equation}\label{barI}
   \bar{I}_{i\leftarrow j}(z)
   = \delta(1-z)\,\delta_{ij} - a_s 
   \left[ {\cal P}_{i\leftarrow j}^{(1)}(z)\,\frac{L_\perp}{2} 
    - {\cal R}_{i\leftarrow j}(z) \right] + {\cal O}(a_s^2)\,.
\end{equation}
The logarithmic piece is proportional to the Dokshitzer{-}Gribov{-}Lipatov{-}Altarelli{-}Parisi (DGLAP) splitting functions ${\cal P}_{i\leftarrow j}^{(1)}$ at one loop. For completeness, these are listed in Appendix \ref{app:b}, together with the remainder functions ${\cal R}_{i\leftarrow j}$. As discussed above, for $q_T \to 0$, we must count $L_\perp\sim \frac{1}{\sqrt{\alpha_s}}$. To achieve uniform accuracy over the entire low $q_T$ region, we must also include the leading logarithmic piece of the two-loop beam functions
\begin{equation}\label{dbarI}
   \Delta\bar{I}_{i\leftarrow j}(z)
   = a_s^2 \left( {\cal D}_{i\leftarrow j}(z) 
   - 2\beta_0\,{\cal P}_{i\leftarrow j}^{(1)}(z) \right)\frac{L_\perp^2}{8} \,,
\end{equation}
where
\begin{equation}\label{Dfun}
  {\cal D}_{i\leftarrow j}(z) = \sum_k  {\cal D}_{i\leftarrow k\leftarrow j}(z) = \sum_{k} \int_z^1 \frac{du}{u} {\cal P}_{i\leftarrow k}^{(1)}(u) \,{\cal P}_{k\leftarrow j}^{(1)}(z/u)\,.
\end{equation}
Sample contributions to  ${\cal P}_{i\leftarrow j}^{(1)}$ and $ {\cal D}_{i\leftarrow k\leftarrow j}$ for quark-induced processes are depicted in Figure~\ref{fig:PandDterms}. The right diagram in the figure shows that the quark flavor before and after radiation can differ in the two-loop terms. Explicit results for all relevant functions are listed in the appendix  in \eqref{BMfct}. 

The complete beam function at NNLL accuracy is thus a second order polynomial in the logarithm $L_\perp$
\begin{equation}\label{Bfunall}
\begin{aligned}
   {\bar{B}}_i(\xi,x_\perp,\mu) 
   &= \sum_j \int_\xi^1 \frac{dz}{z} \bigg[
   \delta(1-z)\,\delta_{ij} - a_s {\cal P}_{i\leftarrow j}^{(1)}(z)\,\frac{L_\perp}{2} 
      + a_s{\cal R}_{i\leftarrow j}(z) \\
   &+ a_s^2 \left( {\cal D}_{i\leftarrow j}(z)- 2\beta_0{\cal P}_{i\leftarrow j}^{(1)}(z) \right)
      \frac{L_\perp^2}{8} \bigg] \, \phi_j(\xi/z,\mu) \\
   &\equiv B^{(0)}_i(\xi,\mu) + a_s \, B^{(1)}_i(\xi,\mu) - a_s \frac{L_\perp}{2} \, B^{(2)}_i(\xi,\mu) \\
   &\quad + a_s^2L_\perp^2\left( - \frac{\beta_0}{4} \, B^{(2)}_i(\xi,\mu)
    + \frac18 \, B^{(3)}_i(\xi,\mu)\right) \,, 
\end{aligned}
\end{equation}
where the coefficients $B^{(m)}_i(\xi,\mu)$ are functions of the renormalization scale $\mu$ and $\xi$, the fraction of the incoming momentum which enters the hard process after the soft and collinear emissions. To be able to work efficiently with these functions, we tabulate them and use a PDF code for their interpolation. 

With the coefficients $B^{(m)}_i(\xi,\mu)$ at hand, the Fourier integral reduces to a set of integrals involving the $n$-th power of a logarithm 
\begin{equation}
{\cal M}_n(Q,\mu,q_T)  =\frac1{4\pi} \int d^2x_{\perp}\,
  e^{-iq_{\perp}\cdot x_{\perp}} \,e^{g_F(\eta_F,L_\perp,a_s)}\, L_\perp^n\,.
  \end{equation}
Since the logarithm $L_\perp$ only depends on $x_T^2 = -x_\perp^2$, we can rewrite $q_{\perp}\cdot x_{\perp} = - x_T \, q_T  \cos \phi$ and integrate over the azimuthal angle $\phi$, which yields
\begin{equation}\label{eq:ident}
  \int_{-\infty}^{\infty} d^2x_{\perp}\,e^{-iq_{\perp}\cdot x_{\perp}}
  = 2\pi \int_0^{+\infty}\!dx_T\,x_T J_0(x_Tq_T) \,.
\end{equation}
Due to the oscillatory nature of the Bessel function $J_0(x_T q_T)$, the numerical convergence is slow. It can be improved by using the identity $J_0(x_Tq_T)=\frac{2}{\pi} \mbox{Im} K_0(-ix_Tq_T)$ and then performing a Wick rotation $x_T\rightarrow ix_T$, which leads to 
\begin{equation}
{\cal M}_n(Q,\mu,q_T) 
= -\frac{1}{\pi}\mbox{Im}\int_0^{+\infty}dx_T\,x_T \,K_0(x_Tq_T)
e^{g_F(\eta_F,L_\perp,a_s)} L_\perp^n \,.
\end{equation}
We compute these integrals on the fly when running our code. Alternatively, one could implement the approximate analytical form developed in \cite{Kang:2017cjk}. 

Expressed in terms of the integrals ${\cal M}_n$ and coefficients $B^{(m)}_i$,  the final form of the Fourier integral, as implemented in our code, is
\begin{equation}\label{eq:four}
\begin{aligned}
  {\cal F}_{ij}(Q&,\mu,q_T,\xi_i,\xi_j)
  = \frac1{4\pi} \int d^2x_{\perp}\,
  e^{-iq_{\perp}\cdot x_{\perp}} \,e^{g_F(\eta_F,L_\perp,a_s)}  {\bar{B}}_i(\xi_1,x_{\perp},\mu)\,{\bar{B}}_j(\xi_1,x_{\perp},\mu) \\
 =& {\cal M}_0(Q,\mu,q_T) \left[ B^{(0)}_i(\xi_1,\mu) B^{(0)}_j(\xi_2,\mu) + a_s B^{(0)}_i(\xi_1,\mu) B^{(1)}_j(\xi_2,\mu) 
   + a_s B^{(1)}_i(\xi_1,\mu) B^{(0)}_j(\xi_2,\mu) \right] \\
   & - \frac{a_s}{2} {\cal M}_1(Q,\mu,q_T) \left[ B^{(0)}_i(\xi_1,\mu) B^{(2)}_j(\xi_2,\mu) + B^{(2)}_i(\xi_1,\mu) B^{(0)}_j(\xi_2,\mu) \right] \\
   & + \frac{a_s^2}{4} {\cal M}_2(Q,\mu,q_T) \Bigg[ - \beta_0 B^{(0)}_i(\xi_1,\mu) B^{(2)}_j(\xi_2,\mu) 
      - \beta_0 B^{(2)}_i(\xi_1,\mu) B^{(0)}_j(\xi_2,\mu) \\
   & \qquad \qquad + \frac{B^{(0)}_i(\xi_1,\mu) B^{(3)}_j(\xi_2,\mu)}{2} + \frac{B^{(3)}_i(\xi_1,\mu) B^{(0)}_j(\xi_2,\mu) }{2}
   + B^{(2)}_i(\xi_1,\mu) B^{(2)}_j(\xi_2,\mu) \Bigg] \,.
\end{aligned}
\end{equation}
Putting together the Fourier part  ${\cal F}_{ij}$ with the RG-evolved hard function ${\cal H}_{ij}$ given in \eqref{eq:resHard} we obtain the resummed cross section \eqref{eq:sigma}. 

\section{Event-based resummation}
\label{sec:autom}

The basic method for the automated computation of the resummed cross section involves the following steps. We first generate events in the Les Houches Event File (LHEF) \cite{Alwall:2006yp} format using the {\sc MadGraph} tree-level event generator. We then use a script written during our earlier work on jet veto resummation \cite{Becher:2014aya} to compute the loop correction for each tree-level event and store this information in the event file. The event files are then processed using our code, which reads in the flavors $i$ and $j$  of the incoming partons, and their momentum fractions $\xi_1$ and $\xi_2$, as well as the $Q^2$ for each event. For a given value of $q_T$, the code then computes the function ${\cal F}_{ij}$ and constructs the RG-evolved hard function $\mathcal{H}_{ij}$ using the loop correction provided with each event and the RG evolution factor. The cross section at a given $q_T$ is then obtained as a weight factor
\begin{equation}\label{eq:reweight}
 w=\left( \frac{\alpha_s(\mu)}{\alpha_s(\mu_{\rm Mad})} \right)^k \,\frac{\mathcal{H}_{ij}(\hat{p}_1, \hat p_2, q_1, ... , q_N,\mu) {\cal F}_{ij}(\xi_1,\xi_2,q_T,\mu)}{\phi_i(\xi_1,\mu_{\rm Mad}) \phi_j(\xi_2,\mu_{\rm Mad})} w^0\,,
\end{equation}
where $w^0$ was the original weight of the tree-level event which was generated with the factorization and renormalization scales set equal to a reference value $\mu_{\rm Mad}$. The denominator is needed to remove the PDFs, which are replaced by the beam functions. The exponent $k$ is the power of $\alpha_s$ of the Born level process. For the quark-induced electroweak vector-boson processes we consider here $k=0$. 

The procedure we just outlined is enough to produce a resummed transverse momentum spectrum for a given process, but we would also like to add experimental cuts on the momenta of the leptons produced in the decays of the vectors bosons and to compute related quantities such as $\phi^*$, which are constructed from lepton momenta. To be able to do so, we will construct a sample of events with different $q_T$, which we then analyze at the end. Before discussing how to generate these events, we must ensure that the hadronic recoil is transmitted to the electroweak final state.

\subsection{Recoil effects}

In the derivation of the factorization formula for the cross section at small transverse momentum $q_T$, one systematically expands in small momentum components. In particular, one drops the small transverse momentum of the partons entering the hard scattering process producing the electroweak bosons. In the factorization formula \eqref{eq:sigma}, the hard partons then have tree-level kinematics with
\begin{equation}
\xi_1 p_1 + \xi_2 p_2 = q \,.
\end{equation}
Expanding away the small transverse momenta is appropriate for the computation of the QCD corrections associated with the large scale $Q^2$. It is also useful because the hard part of the process is then given by the tree-level amplitude $d\sigma^0_{ij}$ times corrections factors, allowing us to generate this part using a tree-level generator. Due to the expansion, momentum is no longer conserved exactly. We now have a mismatch between the electroweak kinematics, which has zero transverse momentum, and the hadronic part, in which the beam functions generate hadronic emissions at a low transverse momentum $q_T$. To correct for this, given a hadronic momentum $p_X^\perp = - q_\perp$  parametrized as
\begin{equation}\label{eq:transMom}
q_\perp^\mu  = ( 0, q_T \cos \phi , q_T \sin\phi, 0)\, ,
\end{equation}
we boost the entire tree-level event such that its total transverse momentum becomes  $q_\perp^\mu$. The total cross section is invariant under this transformation, but the tree-level process now has two incoming partons with small transverse momenta. In our reweighting, we use the momentum fractions of the partons before the boost to determine the momentum fractions $\xi_1$ and $\xi_2$ for the beam functions. Doing so, we again neglect small momentum components but the advantage of proceeding this way is that the electroweak final state has the correct transverse momentum. This gives us access to the transverse-momentum distribution of individual final-state particles. The paper \cite{Catani:2015vma} has discussed different schemes for implementing recoil effects, which differ by terms suppressed by $q_T^2/Q^2$. These power suppressed terms are not captured by the resummation formula, but we will match to the fixed-order results to account for them up to $\mathcal{O}(\alpha_s)$, see below.

\begin{figure}[t!]
\centering
\begin{subfigure}{.5\textwidth}
  \centering
  \includegraphics[width=0.8\linewidth]{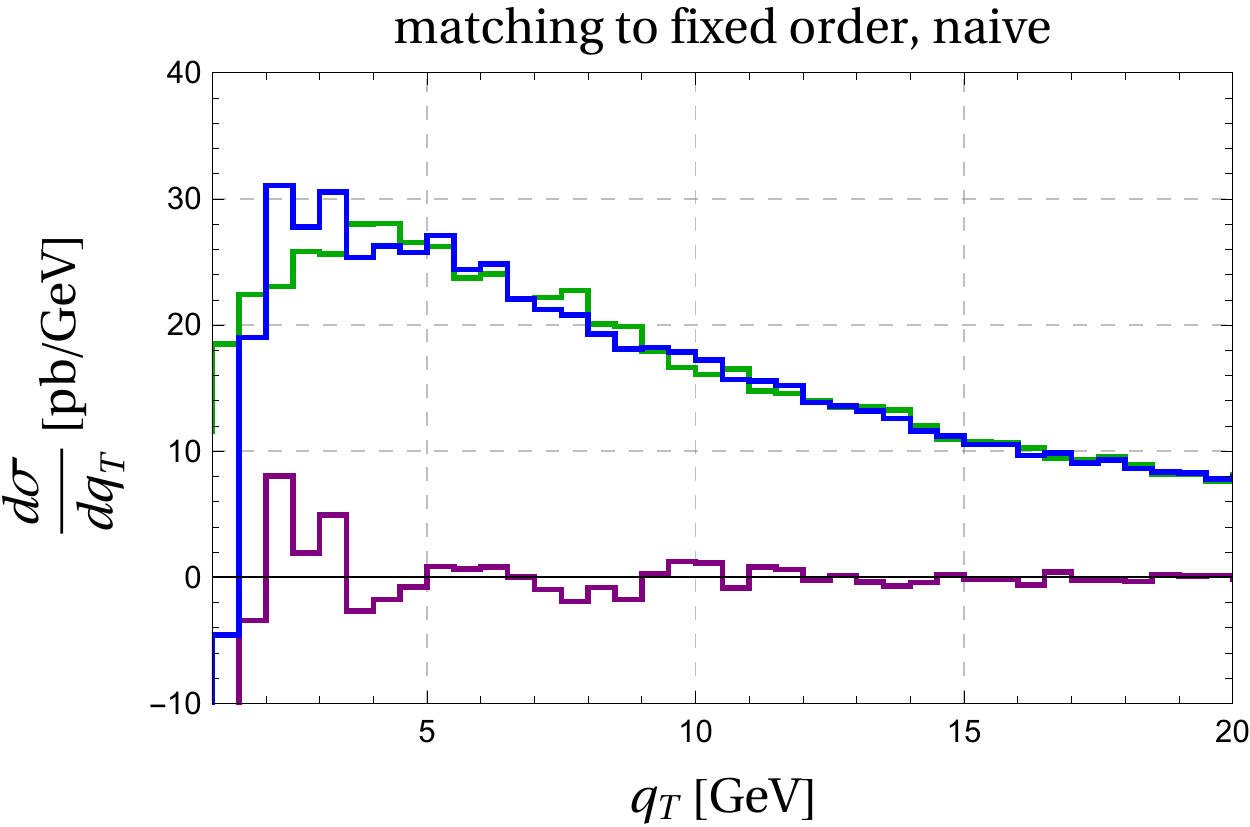} 
\end{subfigure}%
\begin{subfigure}{.5\textwidth}
  \centering
  \includegraphics[width=0.8\linewidth]{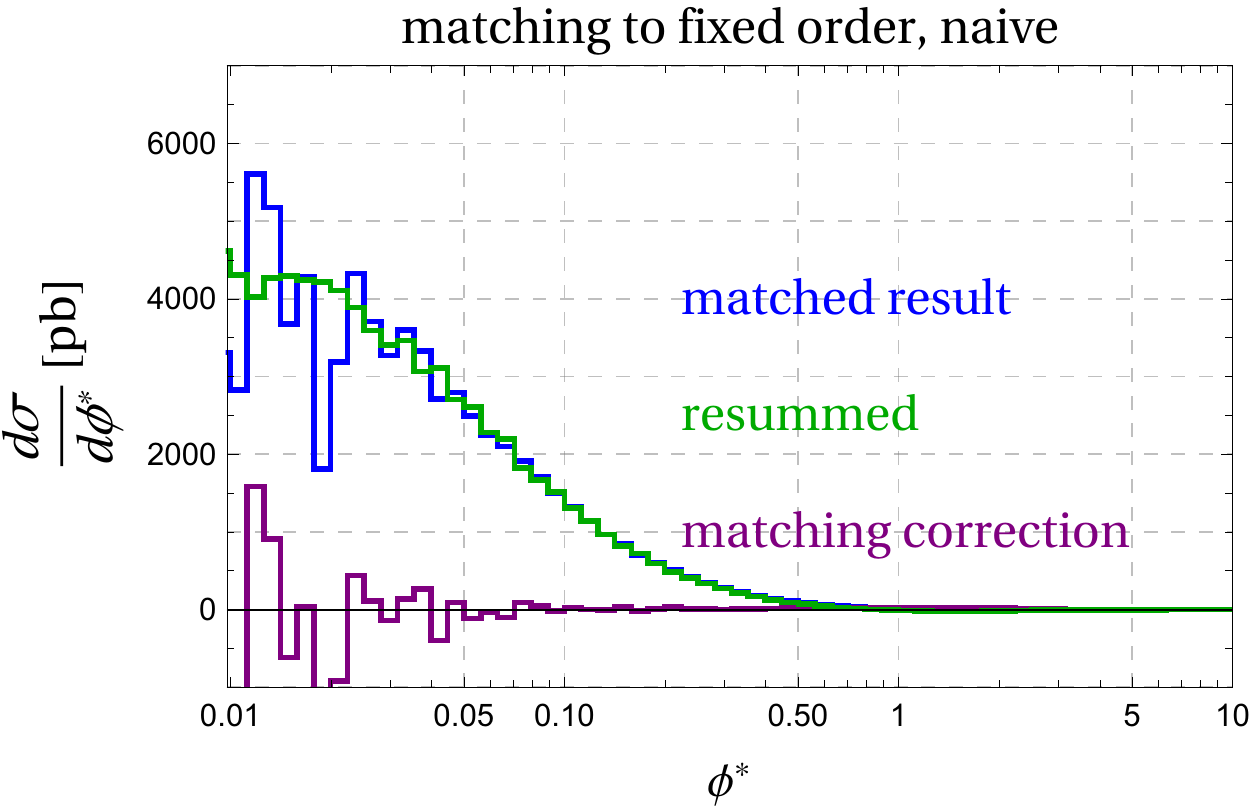}
\end{subfigure}
\caption{Matching correction, resummed and matched result for $q_T$ and $\phi^*$. The numerical noise at small $q_T$ and $\phi^*$ arises due to large cancellations in the naive computation of the matching correction.}
\label{fig:nvmtchd}
\end{figure}

\subsection{Sampling of $q_T$ values}

As indicated above, we want to generate a sample of events with different transverse momenta. The most natural way of doing this, would be to distribute the events according to the cross section, i.e.\ to compute
\begin{equation}
z = \Sigma(q_T) = \int_0^{q_T} d q_T' \frac{1}{\sigma} \frac{d\sigma}{dq_T'}.
\end{equation}
Inverting this relation one obtains $q_T(z)$ and can then use a random number $z\in [0, 1]$ to generate $q_T$ values. Proceeding in this way would yield events with equal weight, but a disadvantage of the procedure is that one would obtain only few events at larger $q_T$ values where the cross section is small. 

In order to have a sample which also covers the region of larger $q_T$ values, we instead generate weighted events by sampling the $q_T$ values uniformly, i.e.\ we generate a random number $z$ and set
\begin{equation} \label{qtval}
q_T = z \, q_{\rm max} \,.
\end{equation}
Imposing a maximum $q_T$ value is necessary in any case because the resummed results for the cross section becomes unphysical at large values $q_T \gtrsim Q$. The value of $q_{\rm max}$ must be large enough to cover the entire region where the resummation is relevant. Choosing $q_{\rm max} \approx Q$ is clearly large enough. Using even larger values would not affect the final result once the matching to fixed order is performed (see Section \ref{sec:matching} below), but would make the event generation inefficient. Writing $\Delta q \coloneqq dq_T/dz = q_{\rm max} $, the cross section integral takes the form
 \begin{equation}
\sigma_{\rm fid} =  \int_{0}^{q_{\rm max}} d q_T  \frac{d\sigma}{dq_T} = \int_0^1 dz\, \Delta q  \frac{d\sigma}{dq_T}\,.
\end{equation}
In a MC evaluation of the above integral with $N$ events, each event thus contributes a weight
\begin{equation}
w = \frac{1}{N} \frac{\Delta q}{\sigma_{\rm fid}}\,  \frac{d\sigma}{dq_T}\, .
\end{equation}
Equivalently, we can assign a cross section
\begin{equation}
\Delta \sigma = \frac{\Delta q}{N}\,  \frac{d\sigma}{dq_T}
\end{equation}
to each event. In the practical implementation, we start with {\sc MadGraph} tree-level events, generate $q_T$ according to \eqref{qtval} and a random angle $\phi\in(0,2\pi)$ to obtain the transverse momentum vector \eqref{eq:transMom}. Then we boost the event in the LHEF as discussed above and compute the event weight using \eqref{eq:reweight}. The boosted momenta and the event weight are then written back into the event file. In a final step, we analyze the resummed events, impose cuts and read out the observable of interest.

\subsection{Matching to fixed order}
\label{sec:matching}

Our resummed result captures logarithms which arise at small transverse momentum but expands away contributions which are suppressed by powers of $q_T^2/Q^2$. At larger transverse momentum these become more and more relevant and should be included. In order to obtain a result which covers all transverse momentum values, we combine
our result with the fixed-order prediction. The labelling of fixed-order results is not uniform in the literature. We will use the term NLO to denote the $\mathcal{O}(\alpha_s)$ result, so that the LO prediction is a $\delta$-function term at $q_T=0$. To avoid double-counting, 
the NLO-expanded NNLL-result must be subtracted from the sum,
\begin{equation}\label{eq:MCorr}
   \frac{d\sigma^{\rm NNLL}}{dq_T} \bigg|_{\text{ matched to NLO}}
   = \frac{d\sigma^{\rm NNLL}}{dq_T} + \underbrace{\frac{d\sigma^{\rm NLO}}{dq_T}
    - \frac{d\sigma^{\rm NNLL}}{dq_T}  \bigg|_{\text{ exp. to NLO}}}_{\text{matching correction $\Delta\sigma$}} \,.
\end{equation}
The first term on the right-hand side of (\ref{eq:MCorr}) is our resummed result, 
the second term the fixed-order NLO result obtained from {\sc MadGraph5\Q{_}aMC@NLO}, 
and the last term the resummed result expanded to NLO. 
The combination of the two latter terms is called the matching correction $\Delta\sigma$.
The NLO-expansion of the resummed result can be obtained using the same reweighting method as for the resummed result; 
the relevant formula is given in Appendix \ref{app:nlo}. The result of this naive matching procedure is shown in Figure~\ref{fig:nvmtchd}.

While formally correct, the matched result (\ref{eq:MCorr}) suffers from two problems. 
First of all, we do not recover the pure fixed-order result, 
even at very large $q_T$, because the resummed result includes higher-order terms in $\alpha_s$. 
Formally they are beyond the accuracy of the computation and can be kept, 
but since they are based on the $q_T\rightarrow 0$ limit, 
they can induce unphysical behavior at large $q_T$. Indeed, naively keeping those terms one ends up with a negative cross section at $q_T \gtrsim Q$. We should therefore switch off the resummation at large transverse momentum.

\begin{figure}[t!]
\centering
  \includegraphics[width=0.4\linewidth]{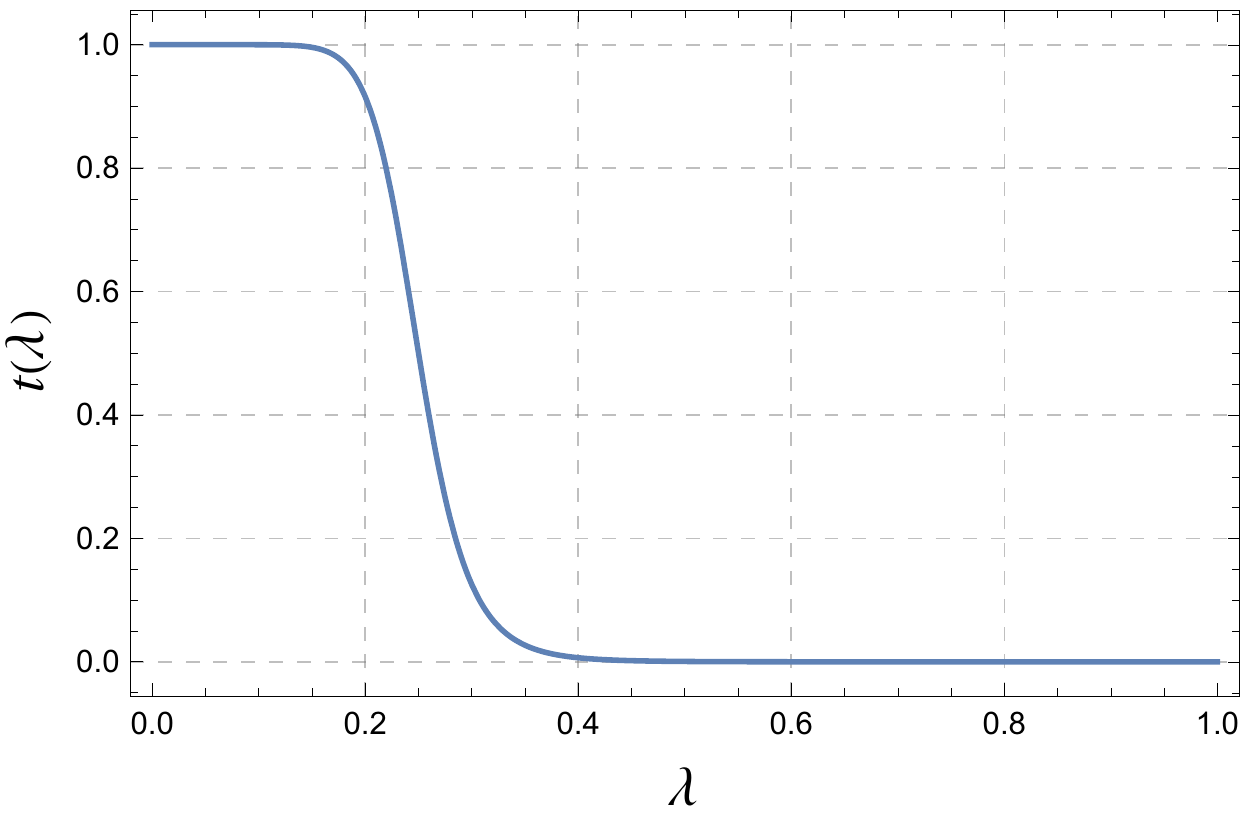} 
\caption{Transition function $t(\lambda)$ used to switch off the resummation.}
\label{fig:trans}
\end{figure}

The second, more immediately visible problem concerns the other end of the spectrum. Both the fixed-order result and the NLO expansion of the resummed result diverge for $q_T\to 0$. The difference goes to zero, but numerically the cancellation is imperfect which leads to large numerical noise that renders the matched result useless for very small $q_T$. The numerical problems are especially visible because the resummed leading-power cross section is Sudakov suppressed for very small $q_T$. Of course the matching correction is not needed in this region and it can even be problematic to include it, because it contains (power suppressed) unresummed large logarithms. In the following, we will improve our matching scheme to solve both of these problems.

To eliminate the numerical noise at small $q_T$, we simply switch off the matching correction for very low $q_T < q_0$, where $q_0$ is a cutoff of the order of a few GeV. The cutoff $q_0$ is chosen large enough to avoid the numerical noise from the incomplete cancellation and small enough that the neglected matching correction, which parametrically scales as $q_0^2/Q^2$, is within the scale uncertainty of the resummed result. Both conditions are fulfilled for the choice $q_0 = 5\,{\rm GeV}$, which we adopt as our default value. 

\begin{figure}[t!]
\centering
\includegraphics[width=1.0\linewidth]{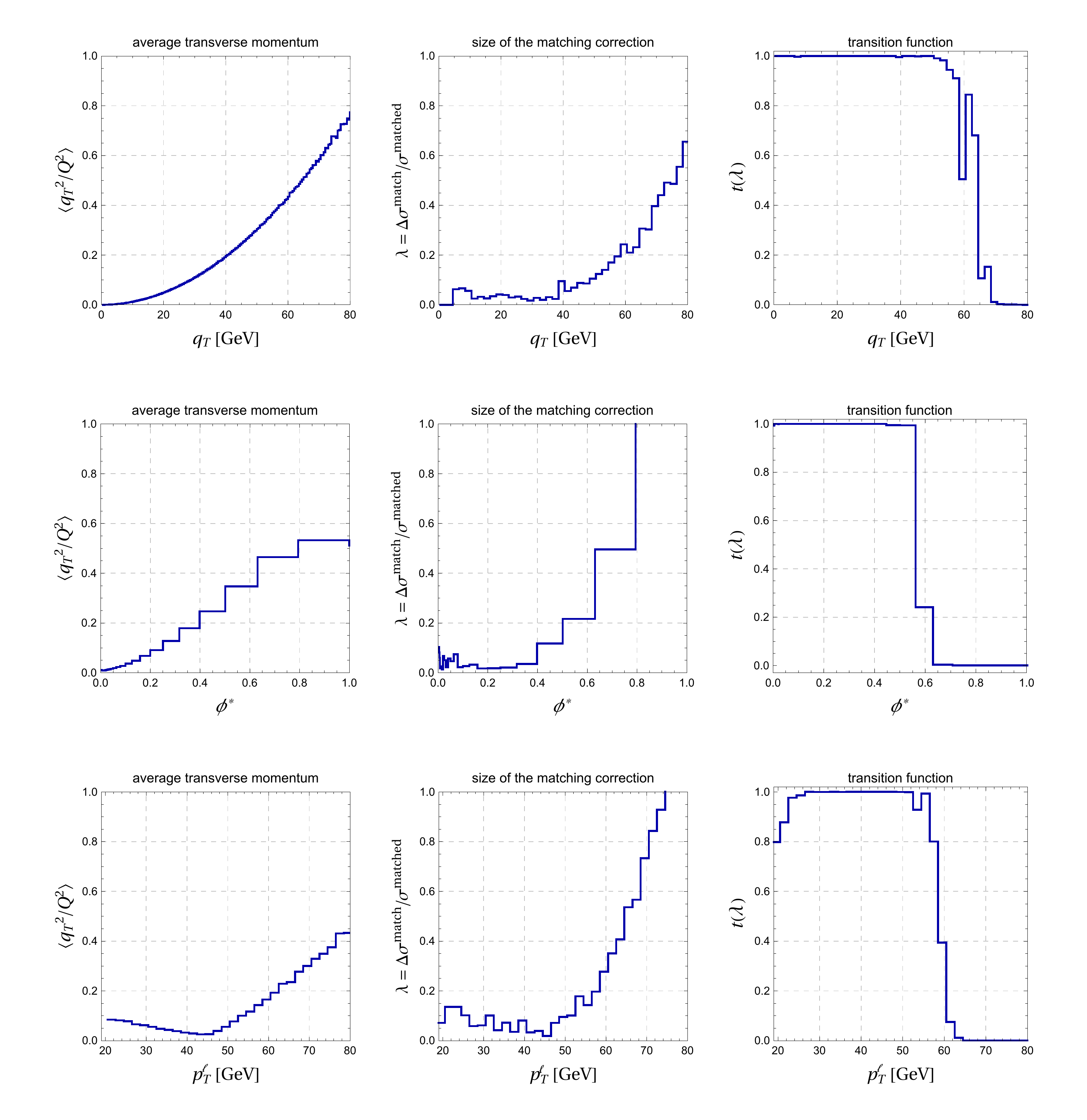}
\caption{Average transverse momentum, size of the matching correction and transition function
for the observables $q_T$, $\phi^*$ and $p_T^\ell$. The expectation values were computed using resummed events with $q_T<Q$.}
\label{fig:transit}
\end{figure}

To switch off the resummation at large $q_T$, we introduce a transition function
\begin{equation}
\label{eq:tlamb}
t(\lambda) := \frac{1}{1+a\, \lambda^{C_i\, b}} \,,
\end{equation}
with $\lambda=\Delta\sigma/\sigma^{\text{matched}}$, where $\Delta\sigma$ is the matching correction and $\sigma^{\text{matched}}$ the naively matched cross section (\ref{eq:MCorr}). We use $a=4$, $b=8$ and $C_i = C_F=4/3$ for the quark induced processes discussed here. The resulting functional form is plotted in Figure~\ref{fig:trans}. The plot shows that we start switching off the resummation when the power suppressed terms amount to 20\% of the result and switch if off completely once they are larger than 40\% of the total cross section. While the value of $\lambda$ is affected by the numerical noise at low transverse momentum, this does not present a problem, since $t(\lambda)$ is equal to one in the region of low $\lambda$, see Figure~\ref{fig:trans}.
 
With the cutoff and the transition function in place, formula \eqref{eq:MCorr} gets replaced by
\begin{equation}\label{eq:impMC}
   \frac{d\sigma^{\rm NNLL}}{dq_T} \bigg|_{\text{ matched to NLO}}
   = t(\lambda)\left(\frac{d\sigma^{\rm NNLL}}{dq_T} + \Delta\sigma \bigg|_{q_T>q_0}\right)
     + \left(1 - t(\lambda)\right)\frac{d\sigma^{\rm NLO}}{dq_T} \,.
\end{equation}
For low values of $q_T$, the function $t(\lambda) = 1$ up to power corrections so that we reproduce expression  \eqref{eq:MCorr} up to the fact that we switch the matching off at very small $q_T< q_0$. To obtain the matching correction $\Delta \sigma$, we evaluate the NLO result for the cross section in {\sc MadGraph5\Q{_}aMC@NLO} for the observable under consideration, imposing $q_T> q_0$. One then subtracts from this the expanded resummed result imposing the same cutoff. For large values of $q_T$, we have $t(\lambda) \to 0$ so that the first term vanishes and we go back to the fixed-order result. 

\begin{figure}[t!]
\centering
\begin{subfigure}{.5\textwidth}
  \centering
  \includegraphics[width=0.8\linewidth]{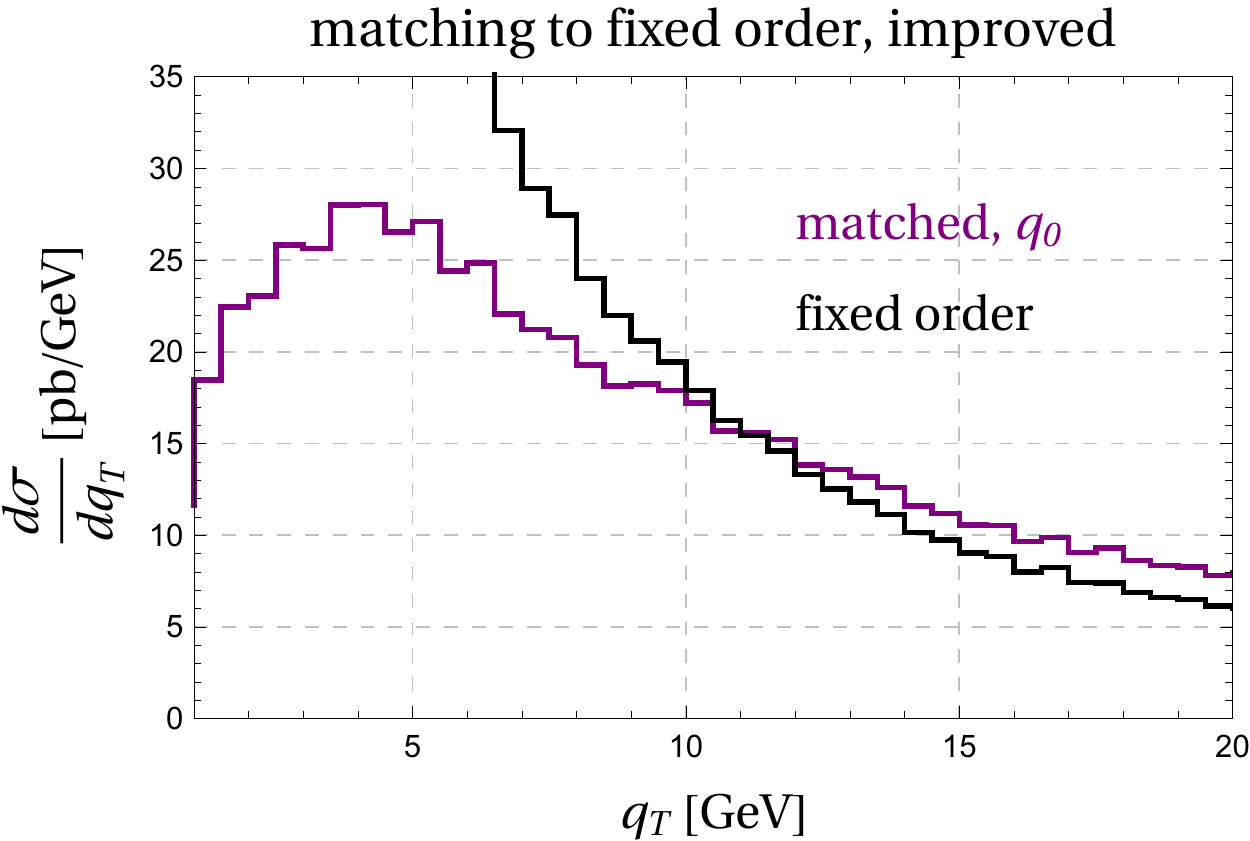} 
\end{subfigure}%
\begin{subfigure}{.5\textwidth}
  \centering
  \includegraphics[width=0.8\linewidth]{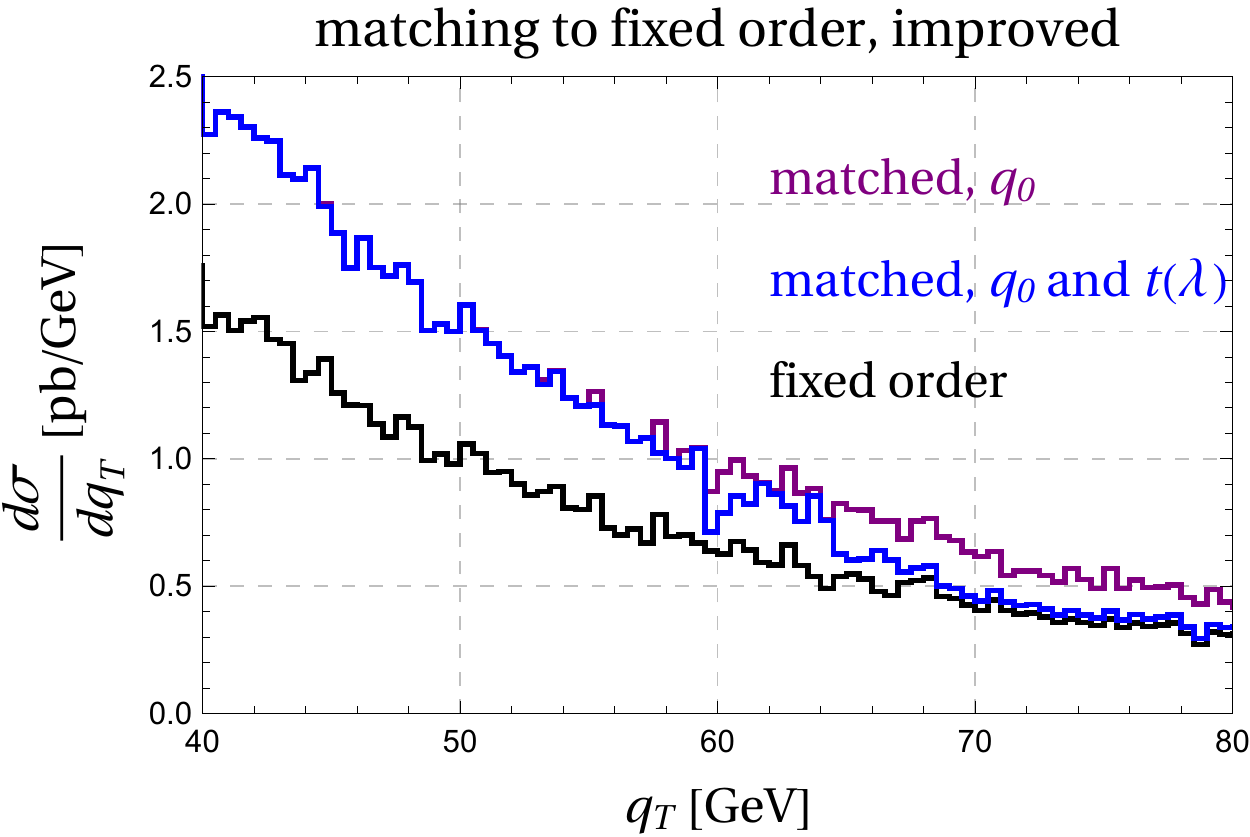}
\end{subfigure}
\caption{Improved matching for $q_T$, according to \eqref{eq:impMC}. The purple curve shows the matching with a cutoff $q_T>q_0$ and the
 blue curve also includes the transition function $t(\lambda)$ which becomes active for $q_T \gtrsim 50$GeV.}
\label{fig:rfmtchdqT}
\end{figure}

There are various other prescriptions to switch off resummation. One can eliminate the logarithms at large $q_T$ by suitably modifying their arguments with power suppressed contributions, or one can use special choices of the renormalization scales to achieve the same.  An advantage of  working with a transition function is that this approach is simple and transparent. In \cite{Catani:2015vma} the transition to fixed order was based on the value of $q_T$. Using instead the size of the power corrections as a measure is useful because it immediately generalizes to other observables such as $\phi^*$ or the lepton momentum distribution. In Figure~\ref{fig:transit} we plot the expectation value of $q_T^2/Q^2$, evaluated with the resummed cross section before matching, the size of the matching corrections and the value of the transition function for $q_T$, $\phi^*$ and $p_T^\ell$, the lepton momentum distribution.  One observes that there is a good correlation between the quantity $\lambda$, which tracks the size of the power corrections, and the expectation value $\langle q_T^2/Q^2 \rangle $ in the region of low transverse momentum. So the power corrections exhibit the expected scaling behavior. 

Some care is required when computing expectation values using the resummed events since the resummed cross section becomes negative at large $q_T \gtrsim Q$ if the matching corrections are not included. We therefore restricted the sample to events with $q_T<Q$ when computing the expectation values for Figure~\ref{fig:transit}. While the expectation value of $q_T^2/Q^2$ and  $\lambda$ display similar behavior, we prefer to use $\lambda$ since it does not require any additional computations beyond the ingredients of \eqref{eq:impMC}. In Figure~\ref{fig:rfmtchdqT}, we show the matched result based on the improved formula \eqref{eq:impMC}. One observes that the numerical noise at small $q_T$ is gone. The plot on the right shows the transition from the resummed result to the fixed-order case which takes place between $q_T$ values of $50-70\, {\rm GeV}$. Our results reduce to the fixed-order predictions for higher values of $q_T$.

\subsection{Implementation}

Let us briefly summarize the relevant steps to obtain a resummed cross section for $p p \to Y + X$, where $Y$ is the electroweak final state under consideration and $X$ the hadronic part of the final state which we assume to have low transverse momentum.
\begin{enumerate}
\item Use {\sc MadGraph5\Q{_}aMC@NLO} to produce tree level events for the process $p p \to Y$, given as an event file in LHEF format.
\item Use the code {\tt virt\_reweighter.py} to compute the virtual correction for each tree-level event. Write this information back into the LHEF.
\item Run our code {\tt qT\_reweighter.py}. This code generates a transverse momentum for each event, boosts the electroweak particles to transmit the recoil and computes the resummed cross section at the given transverse momentum. To this end it assembles the hard function using the virtual correction computed in the previous step, computes the necessary Fourier integrals ${\cal M}_i$ and combines them with the interpolated beam function coefficients $B_f^{(k)}$. The code then writes the boosted vectors and the cross section as a weight back into the event file. In fact, to estimate the uncertainty, the result is computed not only with the default scale choices, but also after varying the scales $\mu$ and $\mu_h$ by a factor of two. Furthermore, we not only compute the resummed cross section, but also its fixed-order expansion to be able to perform the matching. The end result of this step is a statistical ensemble of events with transverse momentum containing different weights for different scale choices.
\item\label{analysis} In the next step, we compute the cross section by analyzing the events for the given observable (such as $q_T$ or $\phi^*$) imposing also the relevant experimental cuts (such as the transverse momentum and rapidity cuts which ATLAS puts on the leptons). At this point, we fill a set of histograms for the observable under consideration, containing the resummed result as well as its NLO expansion for different scale choices.
\item \label{fixed} Next we compute the NLO result using {\sc MadGraph5\Q{_}aMC@NLO} and fill two NLO histograms. One is the full NLO result for the observable under consideration, the other one the NLO result with a cut $q_T>q_0$ needed to perform the matching using \eqref{eq:impMC}. 
\item Finally, we combine the ingredients according to \eqref{eq:impMC} to produce the final resummed and matched predictions, together with their scale variation bands.
\end{enumerate}
The first three steps are observable independent and fully automated, while the analysis part in step \ref{analysis} and the fixed-order computation in step \ref{fixed} need to be set up for the specific observable under consideration. 
Our resummation codes can be obtained upon request and step-by-step instructions on how to use them can be found in \cite{disserationMonika}.

\begin{figure}[t!]
\centering
\begin{tabular}{rrr}
  \includegraphics[width=0.45\textwidth]{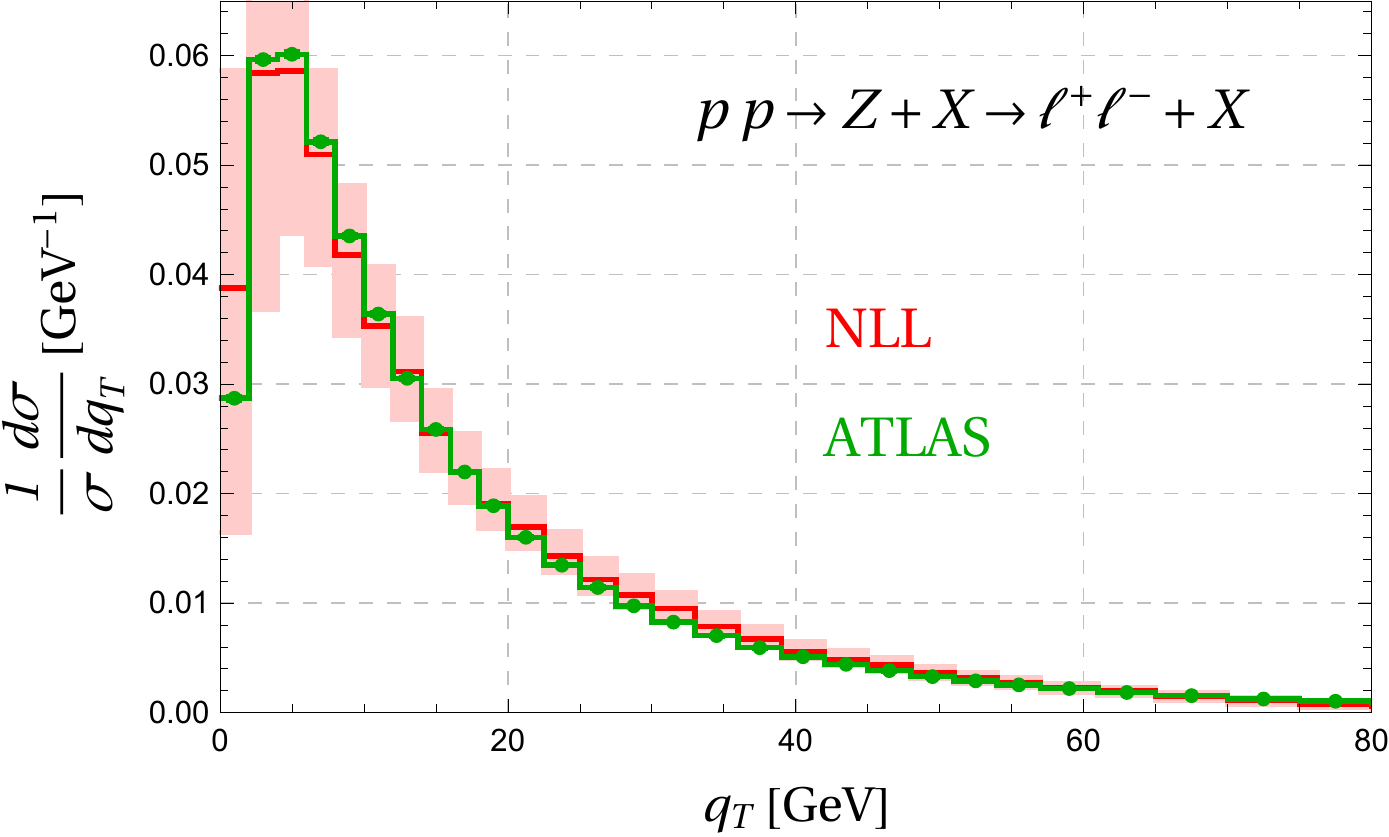} && \includegraphics[width=0.45\textwidth]{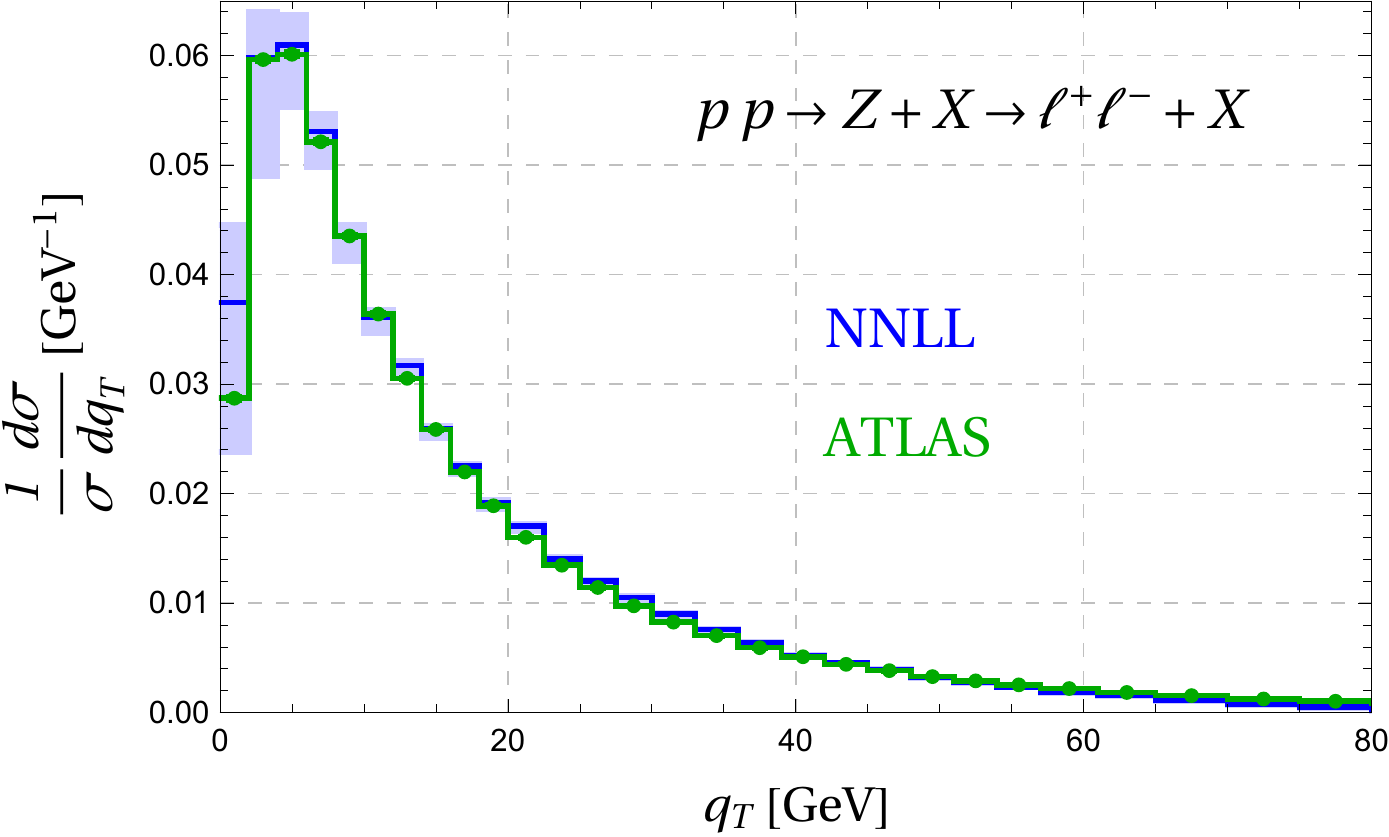}\\
  \includegraphics[width=0.435\textwidth]{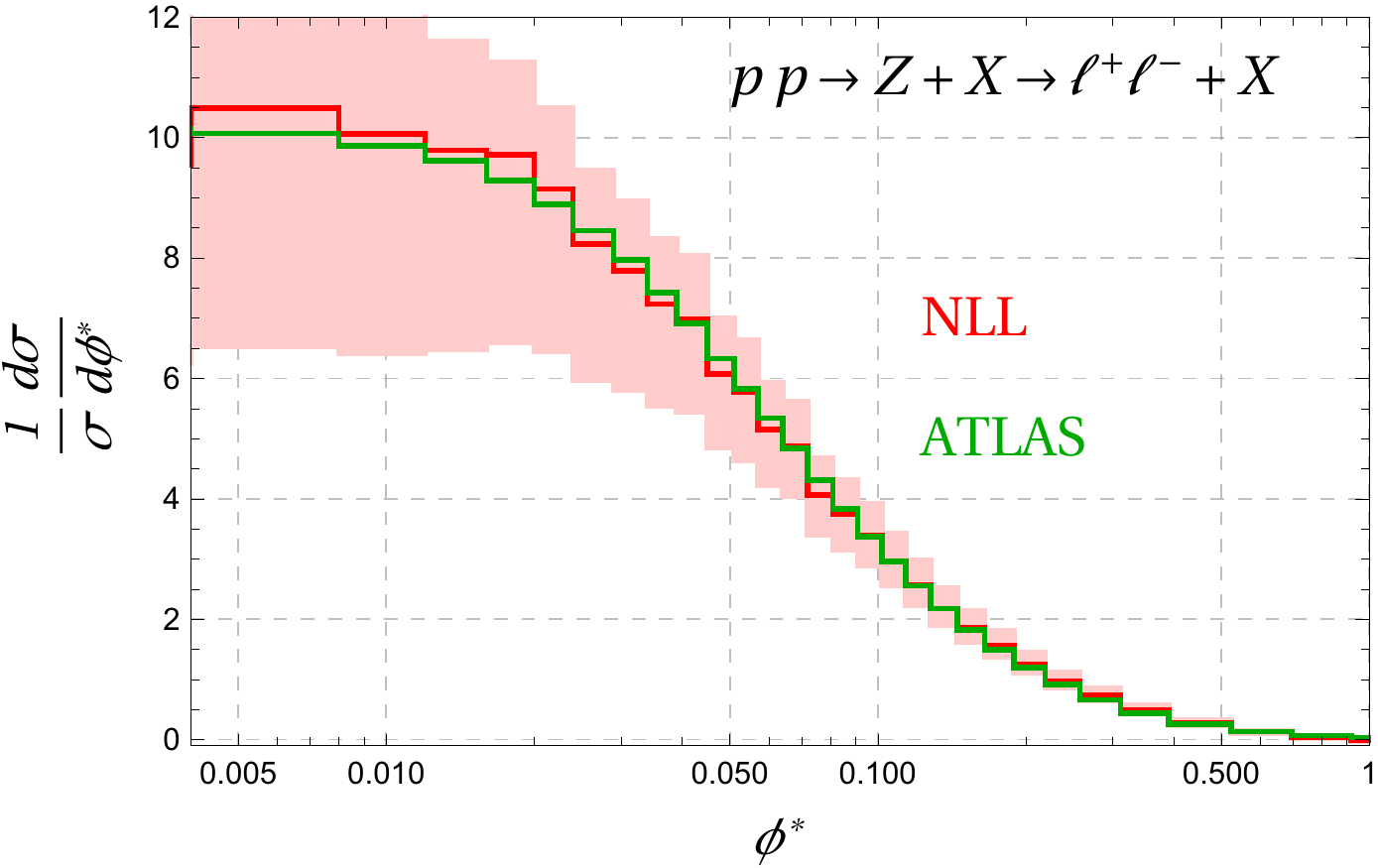}  && \includegraphics[width=0.435\textwidth]{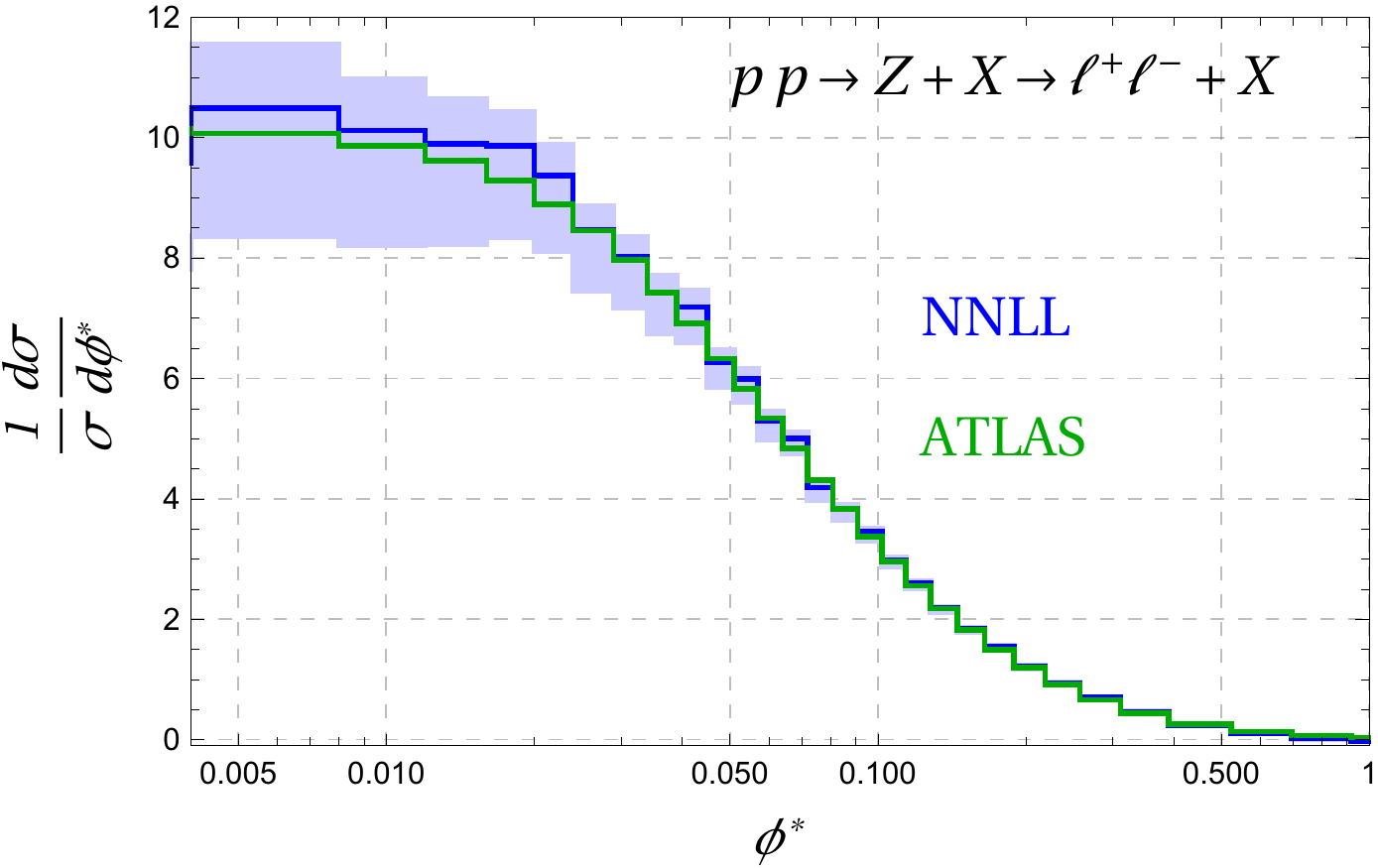}
\end{tabular}
\caption{Comparison of the NLL (red) and NNLL (blue) results for the $q_T$ and $\phi^*$ spectra. The plots show the result without matching.
For visual reference, we also include ATLAS measurements (green points)  \cite{Aad:2015auj}.}
\label{fig:teaser}
\end{figure}

\section{Numerical results}
\label{sec:result}

We now present a few computations made with our code and compare to experimental predictions. Our predictions are based on version 2.6.4 of the {\sc MadGraph5\Q{_}aMC@NLO} framework and unless stated otherwise, we adopt the default parameter values of this code. These include ${M_Z} = 91.188\,{\rm GeV}$, $\alpha_s(M_Z) = 0.118$, $\alpha_{\rm EM} = 1/132.507$, $G_F = 1.16639 \times 10^{-5}\, {\rm GeV}^{-2}$, and the derived quantities $M_W=80.419\,{\rm GeV}$ and $\theta_W=0.490912$. We will work with the MMHT 2014 NLO PDF set with $n_f = 5$ flavors \cite{Harland-Lang:2014zoa}. For the hard scale, we adopt the value $\mu_h=Q$, where the value of $Q$ is set dynamically, on an event-by-event basis. For the low scale, we choose $\mu = q_T +q_*$, where $q_*$ was defined in \eqref{qstar}. The value of $q_*$ is obtained by numerically solving \eqref{qstar} for the value of $Q$ in the event.  In our fixed-order computations and for the matching we set the renormalization and factorization scale to the hard scale, $\mu_f=\mu_r = \mu_h$. To estimate the uncertainties of our computation, we individually vary the scales $\mu$ and $\mu_h$ by a factor of two around their default values and take the envelope of the variations as our scale uncertainty. As expected, the scale bands are driven by the $\mu$ variation at low $q_T$. The $\mu_h$ variation becomes dominant at larger values, when we start to switch off the resummation. 

 \begin{figure}[t!]
\begin{center}
\begin{tabular}{l}
\includegraphics[width=0.45\textwidth]{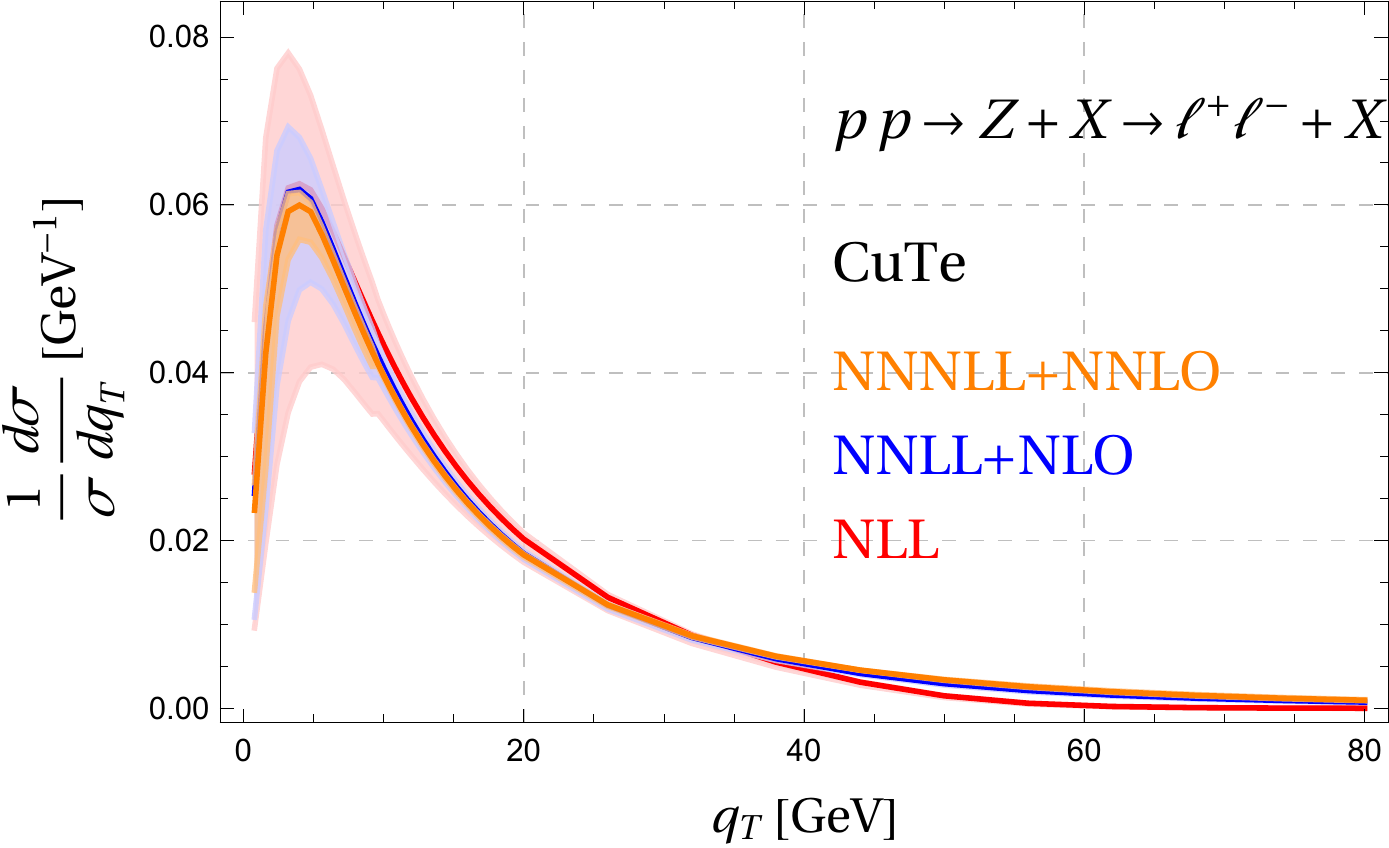}  \\[-0.0cm]
\hspace*{0.5cm}\includegraphics[width=0.428\textwidth]{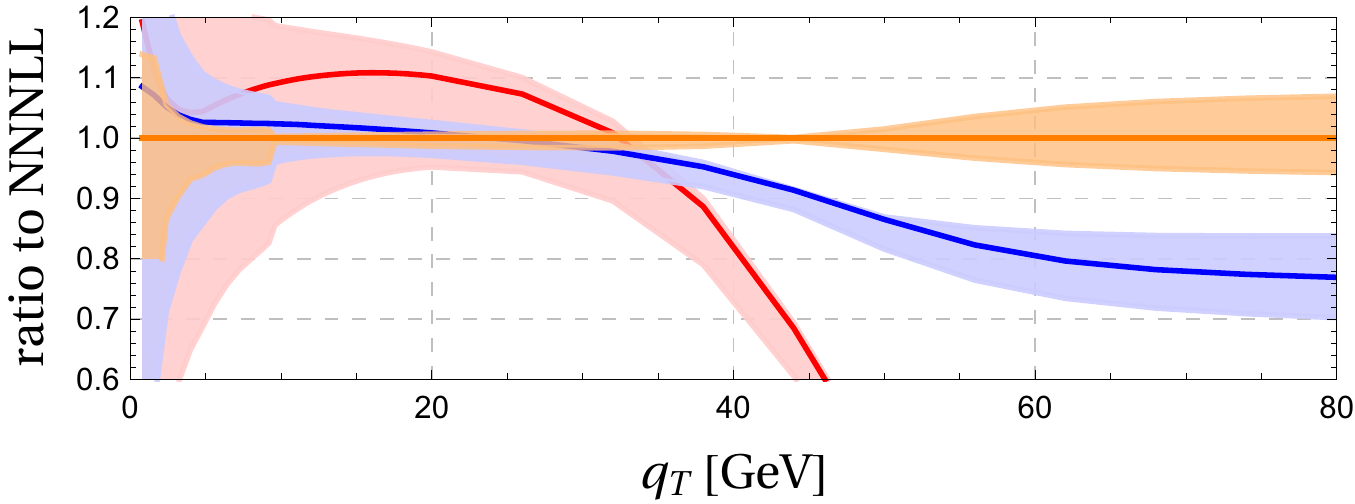} 
\end{tabular}
\end{center}
\vspace{-0.5cm}
\caption{
Resummed $Z$ boson spectrum for $|y|<2.4$ at $\sqrt{s}=8\,{\rm TeV}$ obtained running the {\sc CuTe} code \cite{Becher:2012yn,CuTe}. We normalize each order to its default cross section value in the momentum region shown in the plot and choose $\mu_h = M_Z$.}
\label{fig:CuTe}
\end{figure}

The simplest process we can consider is $Z$ production with decay to leptons. We will compare our results to ATLAS measurements of the $q_T$ spectrum and the related observable $\phi^*$, which is obtained on the basis of angular measurements on the leptons. 
Before confronting experiment, it is interesting to compare NLL resummation to the predictions at NNLL accuracy. The corresponding spectra are shown in Figure~\ref{fig:teaser}. We observe that the scale uncertainties are reduced by about a factor two by going from NLL to NNLL. We also find that NNLL results lie within the NLL uncertainties. Predictions for the inclusive $q_T$ spectrum at the same accuracy and using the same resummation formalism were already obtained using the {\sc CuTe} code \cite{CuTe} in \cite{Becher:2012yn}. We have verified that we reproduce these earlier results if we adopt the same value of the scales and compute to the same order in the improved expansion for $q_T \to 0$. Version~2 of  {\sc CuTe}  includes resummation for the inclusive spectrum up to NNNLL and performs fixed-order matching up to NNLO.\footnote{Version 2.0.2 of {\sc CuTe} incorporates the results for the four-loop cusp anomalous dimension \cite{Moch:2018wjh} and the three-loop anomaly coefficient \cite{Li:2016ctv,Vladimirov:2016dll}. The code thus achieves full NNNLL accuracy.}

 In Figure~\ref{fig:CuTe} we show the spectrum up to this accuracy. While the corrections are small at low $q_T$, the higher-order matching corrections at larger transverse momentum become significant, about 20\%. One also observes that the fixed-order scale bands from varying $\mu=\mu_f=\mu_r$ underestimate the size of the corrections. The value of $\mu_h$ is mainly important for the normalization of the cross section. Choosing $\mu_h^2 = M_Z^2$ gives a relatively low value for the NLL cross section, which then increases as one goes to higher orders. Adopting instead $\mu_h^2 = -M_Z^2$ \cite{Ahrens:2008qu}, the NLL result overshoots and the higher orders give negative corrections. Since the hard function arises as an overall factor, the choice of the hard scale plays only a minor role for the spectrum shown in 
Figure~\ref{fig:CuTe}. In addition to scale variation, {\sc CuTe} provides other options which can be used to estimate uncertainties. One can e.g.~keep the higher-order corrections in the exponent, or expand them out. One can also vary the order of the improved expansion for $q_T \to 0$, which is implemented up to $[\alpha_s(q_*)]^{5/2}$.
 
\begin{figure}[t!]
\begin{center}
\begin{tabular}{rcr}
\includegraphics[width=0.45\textwidth]{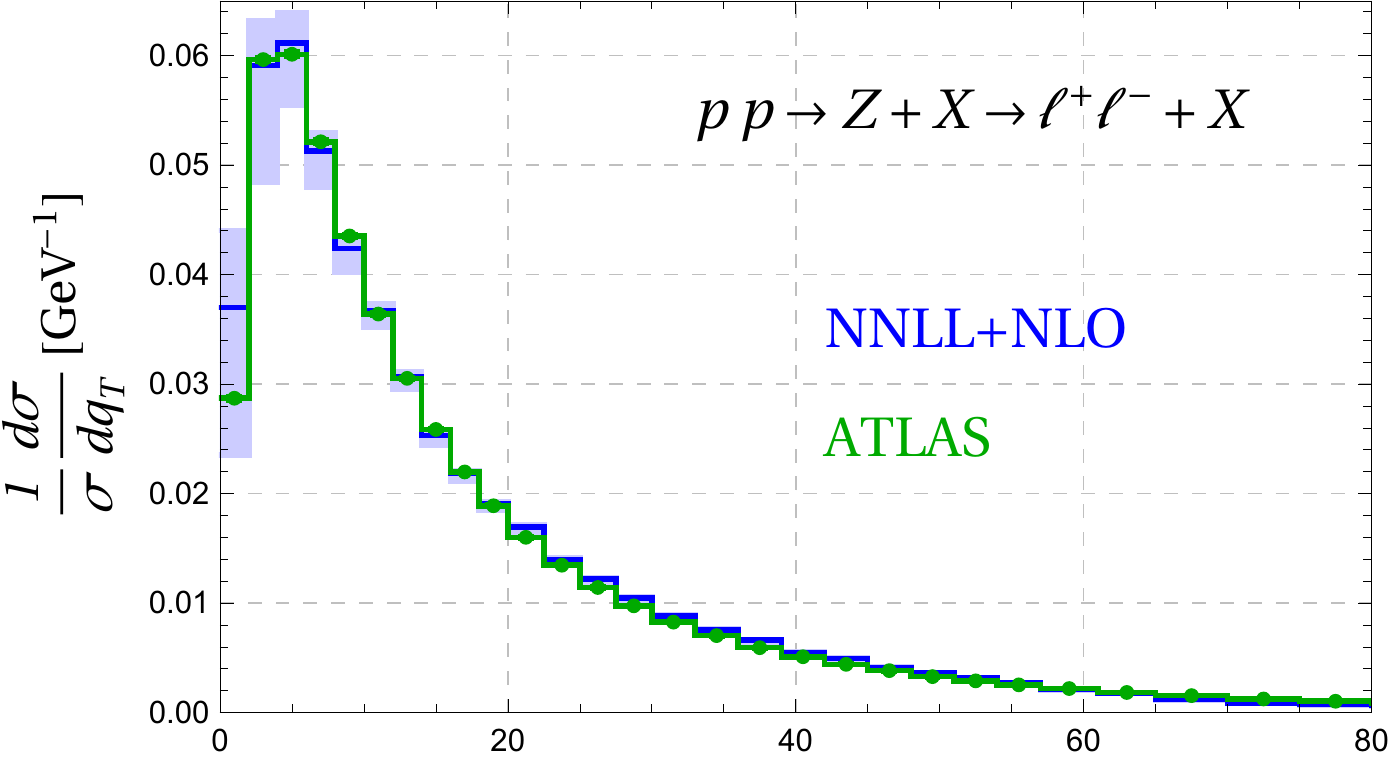} & &
\includegraphics[width=0.44\textwidth]{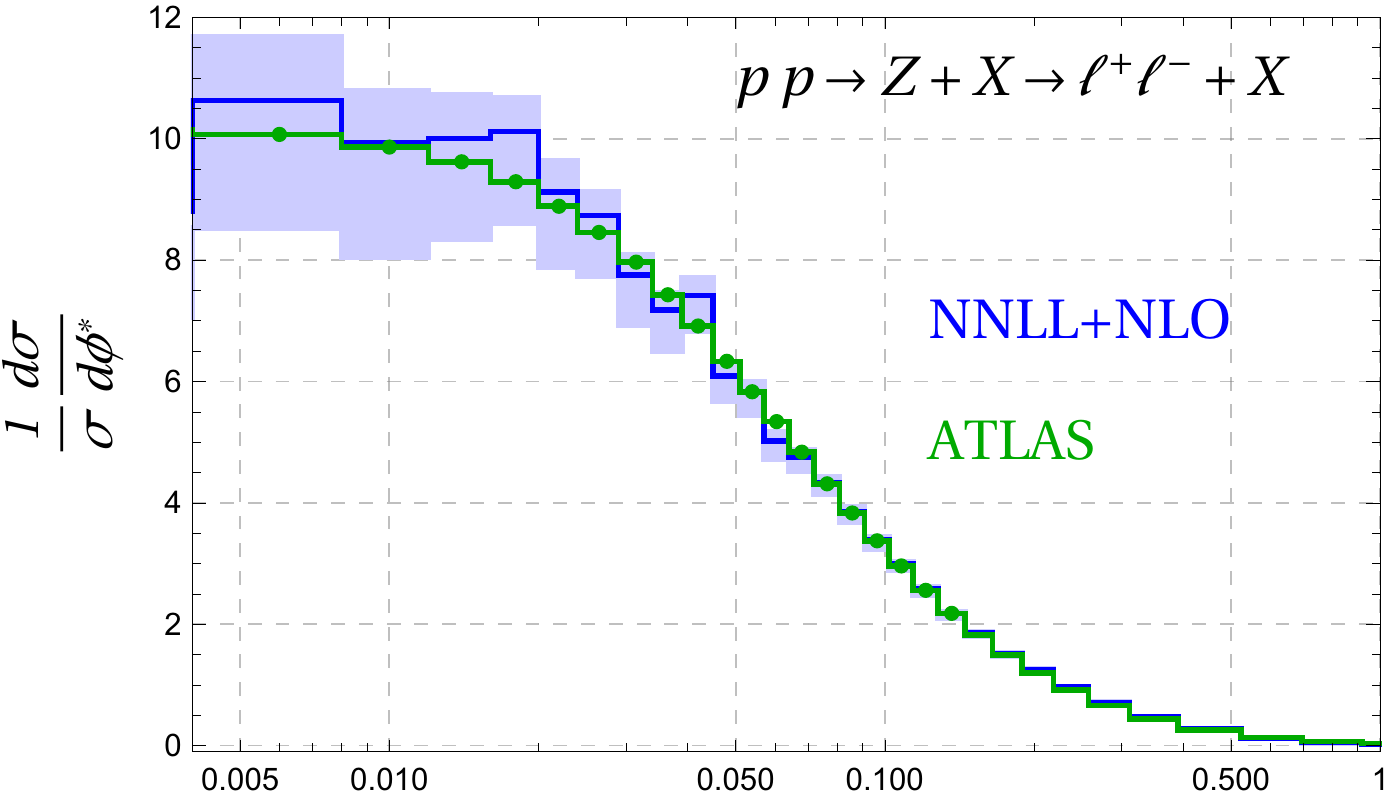} \\[-0.0cm]
\includegraphics[width=0.41\textwidth]{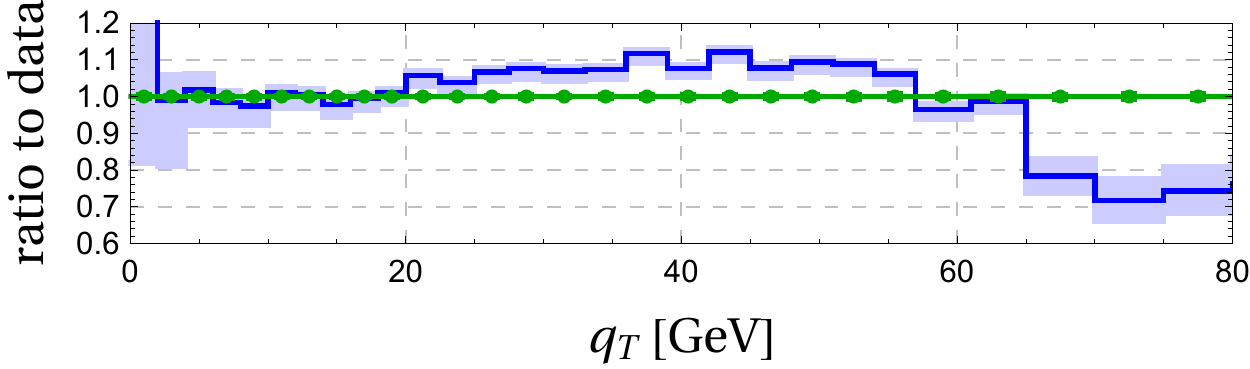} & &
\includegraphics[width=0.42\textwidth]{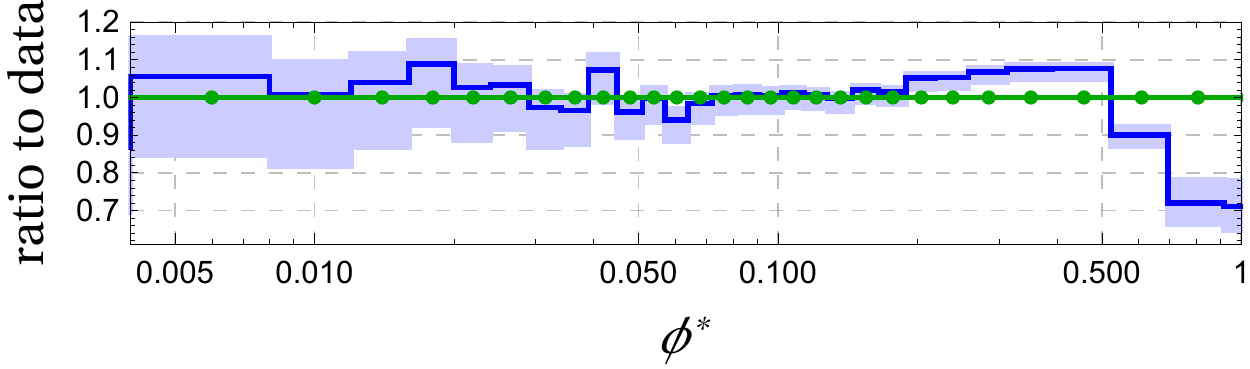} 
\end{tabular}
\end{center}
\vspace{-0.5cm}
\caption{
Comparison of the matched NNLL result to ATLAS data. The experimental uncertainties (green dots) are below $1\%$ and thus invisibly small, the theoretical ones (blue bands) are obtained from scale variation, see text.}
\label{fig:ppZqT}
\end{figure}

What is new compared to {\sc CuTe} is that our code allows us to implement the ATLAS \cite{Aad:2015auj} cuts on the final-state leptons, which are restricted to have $p^\ell_T>  20\,{\rm GeV}$ and pseudorapidity $|\eta| < 2.4$. We focus on the $Z$ resonance region and restrict the invariant mass of the lepton system to the region $66 \,{\rm GeV} < m_{\ell^+\ell^-} < 116\, {\rm GeV}$. In contrast to our earlier work we are able to directly compare our resummed results to the measurement. Furthermore, we can also compute $\phi^*$ since we have access to the full lepton kinematics. The final ingredient for the comparison to ATLAS is the normalization, as our code produces cross sections not spectra. We first compute the fiducial cross section in the region $q_T = 2 - 80\,{\rm GeV}$ from our matched result (using default scale choices) and then divide by this number to get the spectrum. The lower bound at $2\,{\rm GeV}$ is imposed to reduce sensitivity to possible non-perturbative effects. We also normalize the experimental result to the measured cross section in this momentum region. The upper bound was chosen because the unmatched resummed cross section turns negative at higher values of $q_T$ which would lead to unphysical behavior in the unmatched spectra shown in Figure~\ref{fig:teaser}.

In Figure~\ref{fig:ppZqT} we plot our matched results for the $q_T$ and $\phi^*$ spectra, with the lepton cuts imposed by ATLAS \cite{Aad:2015auj}. The agreement is generally quite good, but at intermediate values we overshoot a little bit and our cross section is too small in the fixed order region at large $q_T$. Our fixed-order matching at $\mathcal{O}(\alpha_s)$ only includes the leading term for $q_T \neq 0$ and thus has limited accuracy. The {\sc CuTe} results shown in Figure~\ref{fig:CuTe} show that matching to $\mathcal{O}(\alpha_s^2)$ would bring the cross section into agreement with the data. This is confirmed by \cite{Bizon:2018foh} who match to the known $\mathcal{O}(\alpha_s^3)$ result \cite{Gehrmann-DeRidder:2016jns}  and obtain  a result which nicely agrees  with the experimental data. In reference \cite{Bizon:2018foh} the resummation is performed up to NNNLL, which leads to an excellent description of the data over the entire momentum range. In the context of the fixed-order computation, let us mention that in the matching scheme \eqref{eq:impMC} with a cutoff $q_0$ on the matching corrections, we could extend the matching with some effort to $\mathcal{O}(\alpha_s^2)$. To do so, one would use the {\sc MadGraph5\Q{_}aMC@NLO} to perform a NLO computation of $Z + j$ with $p_T^j > q_0$  and also expand the resummed results one order higher in $\alpha_s$ to extract $\Delta \sigma$. 

\begin{figure}[t!]
\begin{center}
 \includegraphics[width=0.45\textwidth]{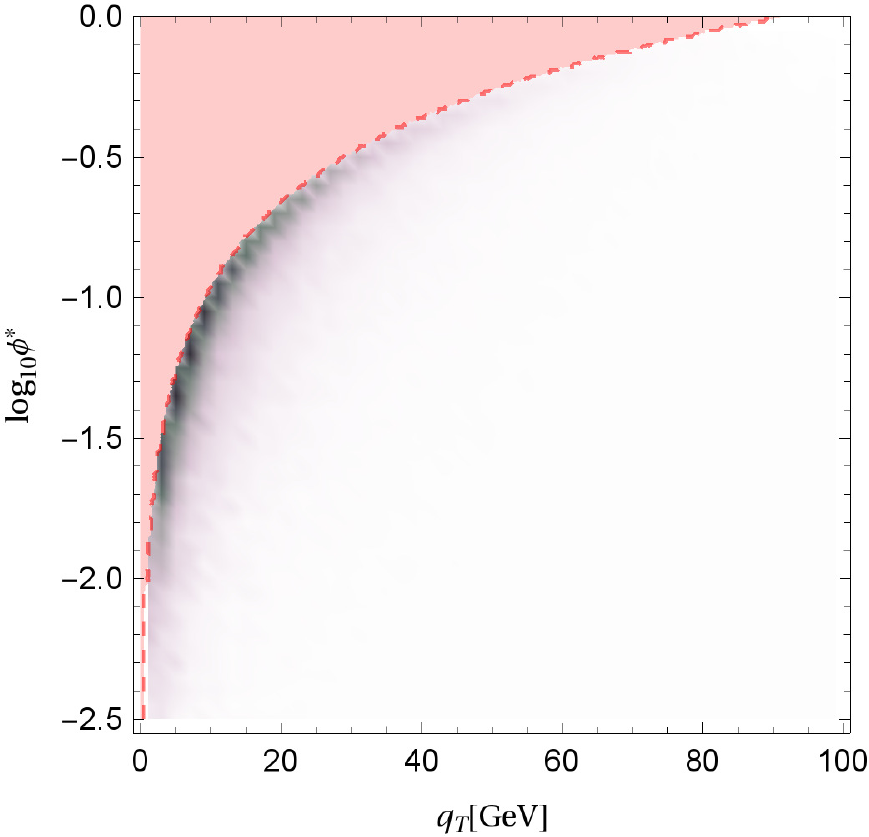} 
\end{center}
\vspace{-0.5cm}
\caption{\label{plCorr} The double differential cross section in $q_T$ and $\log_{10} \phi^*$. The dashed red line corresponds to $ \phi^* = q_T/M_Z$,   the maximum achievable value of  $\phi^*$ for a given $q_T$. In the red region above the dashed line, the cross section vanishes. Dark areas in the density plot correspond to large cross section. Most of the cross section arises from values of $\phi^*$ close to the kinematic boundary. }
\end{figure}

As discussed in the introduction, the variable $\phi^*$ was constructed as an alternative to $q_T$, as it can be measured more precisely. To illustrate their correlation, we show  in Figure~\ref{plCorr} a density plot of the cross section in $q_T$ and $\log_{10} \phi^*$. For a given $q_T$, there is a maximum possible value of $\phi^*$, which is obtained when the two leptons are produced at $\Delta \eta=0$. Determining the minimum $\Delta\phi$ and inserting it into the definition \eqref{phiDef}, one finds that $\phi^*_{\rm max} = q_T/Q$. The corresponding relation (for $Q=M_Z$) is shown as a dashed red line in Figure~\ref{plCorr} and the red area above the line is kinematically excluded. The largest cross section is found near the maximum possible value of $\phi^*$ which demonstrates the close correlation among the two observables. In \cite{Banfi:2009dy} it was observed that the logarithms in the $\phi^*$ distribution which arise at NLO can be obtained from the one in the transverse momentum spectrum by the substitution $q_T/M_Z \to 2\phi^*$, in agreement with our findings.

\begin{figure}[t!]
\begin{center}
\begin{tabular}{cc}
\includegraphics[width=0.45\textwidth]{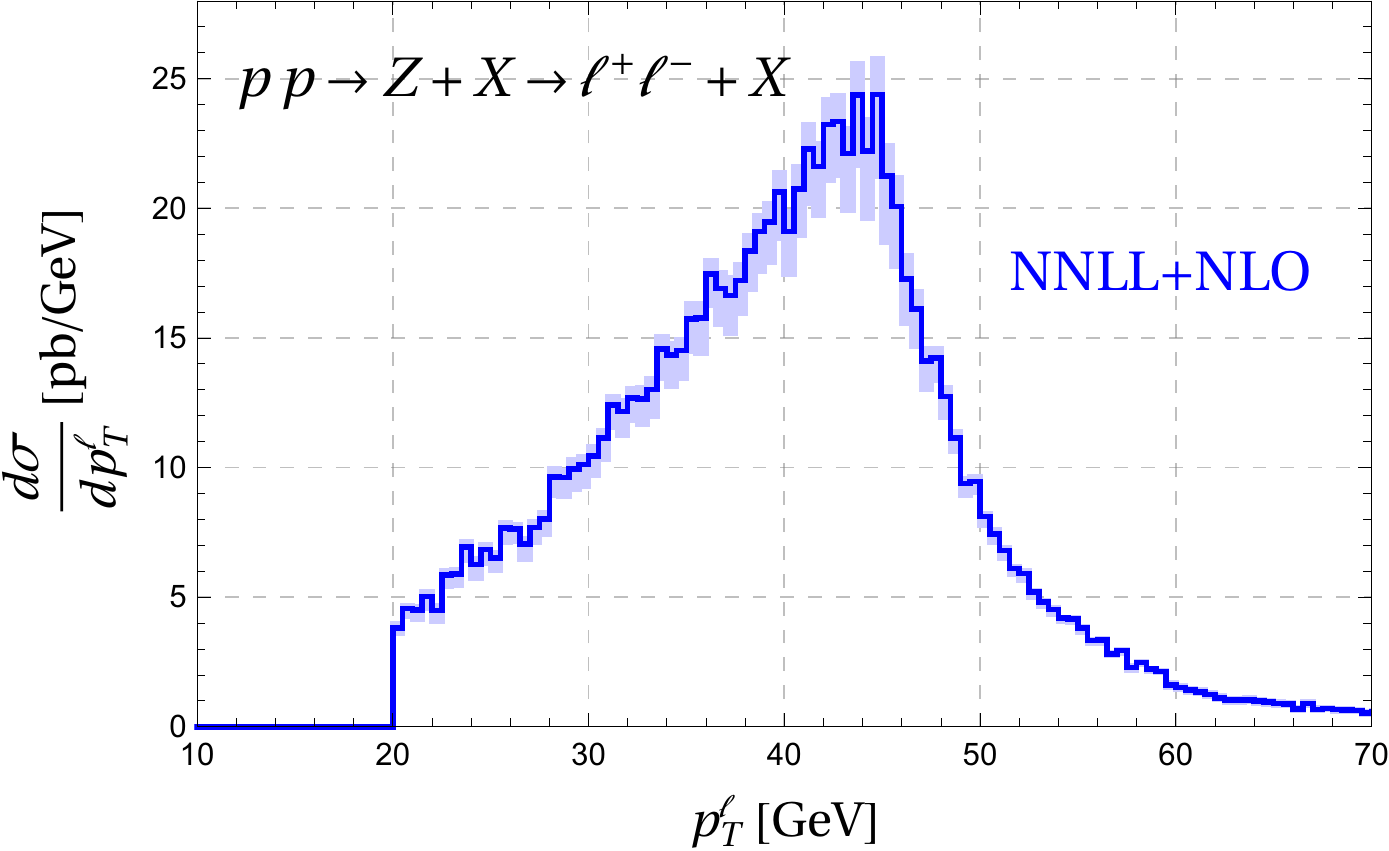} & \includegraphics[width=0.455\textwidth]{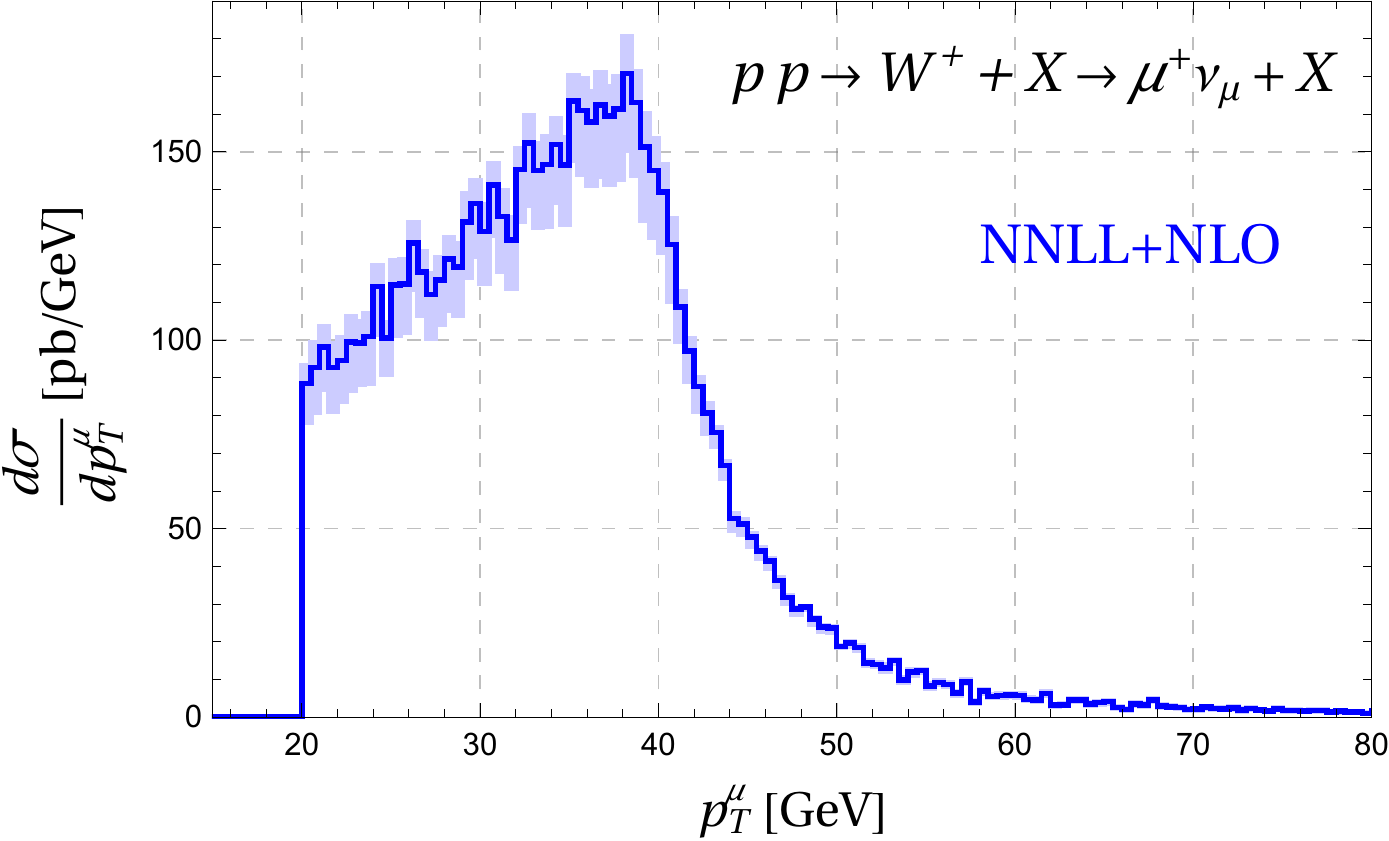} 
\end{tabular}
\end{center}
\vspace{-0.5cm}
\caption{\label{plAbspTel}
Matched NNLL result for lepton momenta. We impose $p_T^\ell> 20\, {\rm GeV}$ and $|\eta_\ell|<2.4$. Left: Lepton transverse momentum in $Z$ boson production at $\sqrt{s} = 8\,{\rm TeV}$. Right: The $\mu^+$ transverse momentum in $W^+$ production at $\sqrt{s} = 7\,{\rm TeV}$. The bands show the scale uncertainties.}
\end{figure}

Since it is interesting in the context of the $W$-mass determination, we also show in Figure~\ref{plAbspTel} the matched result for the lepton transverse momentum distribution in $Z$ production, imposing again the ATLAS \cite{Aad:2015auj} cuts as described above. Due to the lepton transverse momentum cut, this distribution starts at $p_T^\ell = 20\,{\rm GeV}$. For this observable, resummation effects are especially important near the endpoint of the tree-level result at half of the mass of the produced boson. Indeed, for $p_T^\ell \approx M_V/2$ the distribution is dominated by low-$q_T$ events, while the matching becomes important at higher values of $p_T^\ell$, see Figure~\ref{fig:transit}. The lepton momentum spectrum is much easier to measure than the transverse momentum of the weak boson, especially for the $W$ where one has to reconstruct the missing energy to obtain the boson momentum. To our knowledge no LHC measurements were presented for $p_T^\ell$, so that we cannot directly compare to data. The right-hand plot in Figure~\ref{plAbspTel} shows the charged-lepton momentum distribution in $W^+$ production at $\sqrt{s}=7\,{\rm TeV}$ as in \cite{Aaboud:2017svj}, imposing the same cuts on the charged lepton as in the $Z$ boson case.

\begin{figure}[t!]
\begin{center}
\begin{tabular}{cc}
\raisebox{0.002\textwidth}{\includegraphics[width=0.45\textwidth]{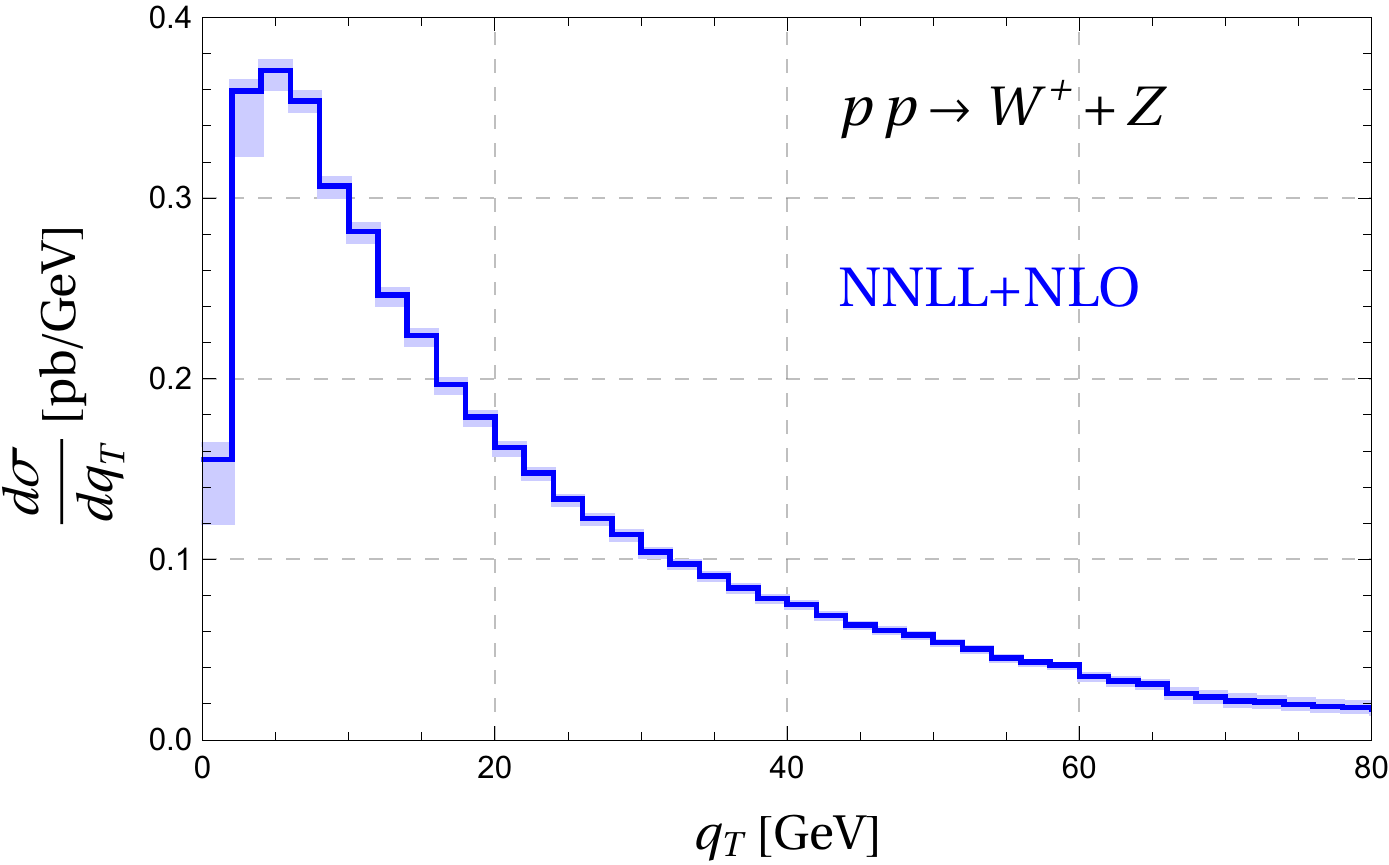}} & \includegraphics[width=0.465\textwidth]{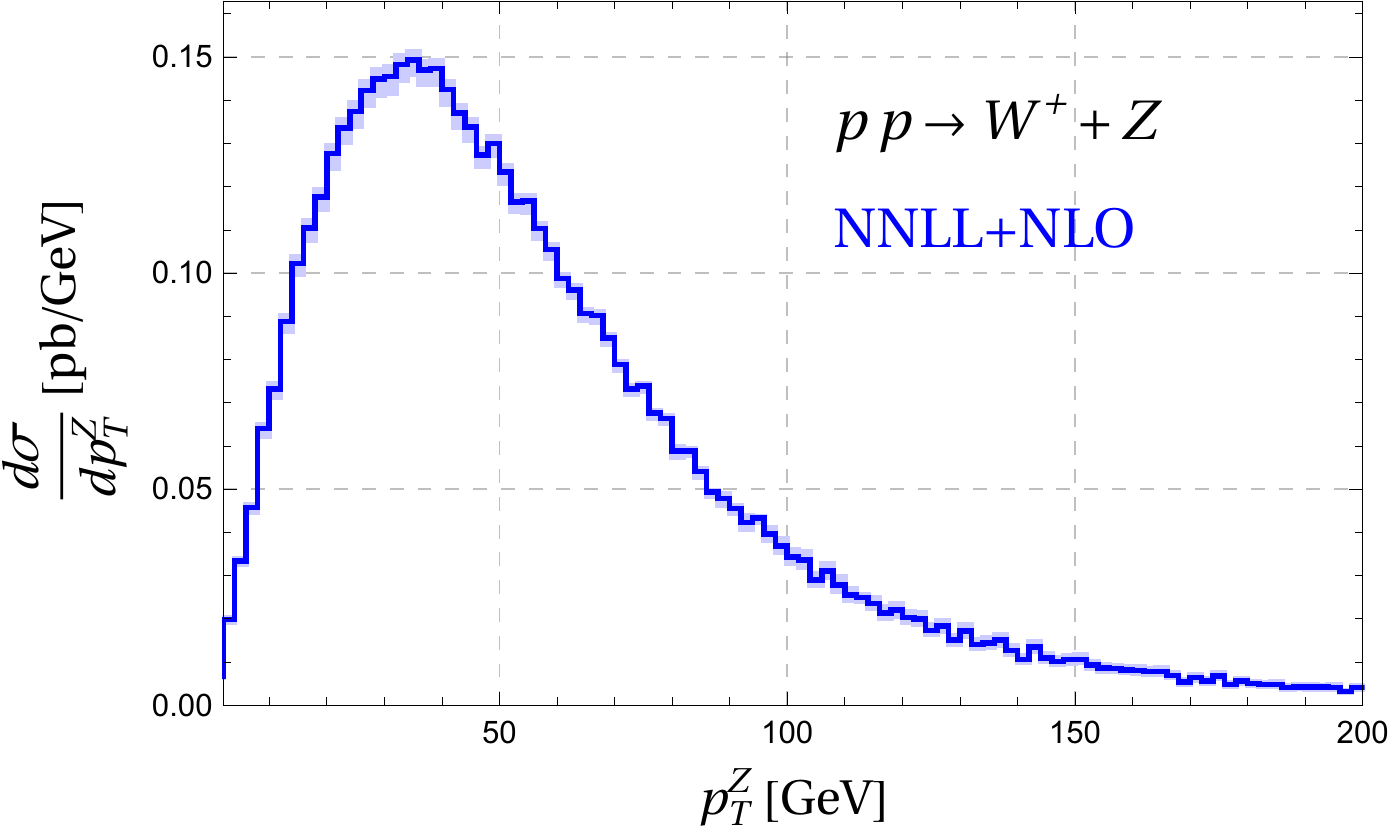} 
\end{tabular}
\end{center}
\vspace{-0.5cm}
\caption{\label{plWpZ}
Matched cross sections for $W^{+}Z$ production at $\sqrt{s} = 7\,{\rm TeV}$. The left plot shows the total transverse momentum $q_T$, the right one  $p^Z_T$, the transverse momentum of the $Z$ boson. The bands show the scale uncertainties. The only cut we apply is on the rapidity of the diboson system, which we restrict to $|y|<2.4$.}
\end{figure}

As discussed earlier, predictions for $q_T$ and for $\phi^*$ at NNLL accuracy have been presented before and in the recent paper \cite{Bizon:2018foh} even NNNLL results were given. The advantage of our implementation is its flexibility to perform transverse momentum resummation for arbitrary massive electroweak final states. As a first example, we have performed resummation for $W^{\pm}Z$  production.  The transverse momentum spectrum of $Z$ bosons in $W^{\pm}Z$ production was measured in \cite{Aad:2012twa} and used to put limits on anomalous triple gauge boson couplings. In Figure~\ref{plWpZ} we show results for the diboson as well as the $Z$ boson transverse momentum spectrum. Resummation for diboson production has been studied earlier in the papers \cite{Grazzini:2005vw, Frederix:2008vb,Balazs:1998bm,Wang:2013qua,Grazzini:2017mhc}.

\section{Conclusions}
\label{sec:concl}

An important benefit of factorization --- obtained using effective field theory or by other means --- is universality: the same low-energy matrix elements typically arise in many different processes. The prime example in a collider context is provided by the parton distribution functions which capture the low-energy physics of arbitrary hard-scattering processes. The same universality is present for electroweak boson production at low transverse momentum. The accompanying QCD radiation is process independent and described by a Fourier convolution of two beam functions. In this paper, we have made use of this universality to automate transverse momentum resummation for arbitrary electroweak final states. To this end we have used {\sc MadGraph5\Q{_}aMC@NLO} to generate the hard process together with its virtual corrections. The universal beam functions are tabulated similar to PDFs and then added to the hard process from the event generator using reweighting techniques. Our results automate the resummation to NNLL as well as the $\mathcal{O}(\alpha_s)$ matching to fixed order and allow us to compute cross sections for arbitrary electroweak production processes with leptonic decays. Our event-based framework allows us to impose experimental cuts on the leptons which arise in the decay of the electroweak bosons and to have access to a variety of variables such as $\phi^*$.

There are a number of additions and improvements which could be incorporated into our framework in the future. First of all, it would be useful to also implement gluon-induced processes up to NNLL so that one could also study Higgs-boson production and associated processes. This would also allow one to study gluon-induced diboson production, which, although loop suppressed, can be sizeable due to the large gluon PDFs. Another improvement would be to extend the fixed-order matching at $q_T > 0$ to $\mathcal{O}(\alpha_s^2)$ by implementing the fixed-order expansion of the resummed result to one order higher. Finally, one could extend the resummation to NNNLL. However, this would require the two-loop hard functions, which would have to be supplied by hand for the cases where they are known. For diboson production, these functions were computed in \cite{Caola:2015ila,Gehrmann:2015ora,vonManteuffel:2015msa} and are provided in numerical form through the {\sc VVamp} code \cite{vvamp}.

In the present paper, we have compared our results to the $q_T$ and $\phi^*$ distributions for $Z$ production measured by ATLAS and we have also computed sample diboson observables to demonstrate that our method also works for more complicated final states. In the future, it would be interesting to study diboson processes in more detail. Their sensitivity to new physics has been pointed out a long time ago \cite{Hagiwara:1986vm,Hagiwara:1989mx} and they are measured increasingly precisely at the LHC \cite{Khachatryan:2015sga,Aad:2016ett,Aaboud:2017oem,Aaboud:2017rwm,Sirunyan:2017zjc, Sirunyan:2019bez,Aaboud:2019lxo}. In this context our method would be quite useful. For example, it was proposed to impose a cut on $q_T$ in the process $pp \to W Z \to \ell \nu \bar{\ell}' \ell'$ to increase the sensitivity to new physics  \cite{Franceschini:2017xkh}. This specific cut is imposed to preserve the amplitude zero present in the Standard Model \cite{Baur:1994ia}, but more generally these types of cuts are useful in order to prevent radiative corrections from washing out the operator structure one wants to probe with the measurements. Such a cut of course leads to Sudakov logarithms which can be resummed using our method. Since our code works with any model  implemented into {\sc MadGraph5\Q{_}aMC@NLO}, we can perform this resummation also for new-physics models or in Standard Model Effective Field Theory, which parametrizes the possible deviations in a model independent manner. The Feynman rules for this effective theory have been automated in a way suitable for the {\sc MadGraph5\Q{_}aMC@NLO} framework \cite{Dedes:2017zog,Brivio:2017btx,Dedes:2019uzs}. We look forward to applications of our framework in the Standard Model and beyond!

\vspace{0.2cm}
{\em Acknowledgments:\/}
This research is supported by the Swiss National Science Foundation (SNF) under grants 200020\_182038 and 200020\_165786. We thank Stephan Caspar, Lorena Rothen and Ding Yu Shao for discussions and Marcel Balsiger for beta testing our code and a thorough reading of the manuscript.

\begin{appendix}

\section{Ingredients of the cross section at NNLL}

\renewcommand{\theequation}{A\arabic{equation}}
\setcounter{equation}{0}

For completeness we provide here all the ingredients for the resummed cross section as well as its fixed-order expansion needed to perform the matching.

\subsection{Evolution of the hard function}
\label{app:c}

The RG evolution factor $U(Q^2,\mu_h,\mu)$ in \eqref{eq:resHard} needed to evaluate the hard function at a low scale
has the form 
\begin{equation}\label{Umat}
U (Q^2, \mu_h,\mu) = e ^ {4 C_i S(\mu_h,\mu) - 4 a_{\gamma_i}(\mu_h,\mu))}
\left(\frac{Q^2}{\mu_h^2}\right)^{-2 C_i a_{\Gamma}(\mu_h,\mu)} \,,
\end{equation}
where $C_q = C_F$ for quark-induced processes and $C_g = C_A$ for the gluon case.  The exponent of the evolution factor involves the single-logarithmic functions
\begin{equation}\label{agamma}
a_{\gamma_i}(\mu_h,\mu) 
= \frac{\gamma^i_0}{2\beta_0}\left[\ln r + \left(\frac{\gamma^i_1}{\gamma^i_0}-\frac{\beta_1}{\beta_0}\right)
\frac{\alpha_s(\mu) - \alpha_s(\mu_h)}{4\pi} + \dots \right] \,,
\end{equation}
with $r=\alpha_s(\mu)/\alpha_s(\mu_h)$, and $a_{\Gamma}(\mu_h,\mu)$, which is given by the same expression after replacing the anomalous dimension by the cusp anomalous dimension, $\gamma^i_n\to  \Gamma_n $. The anomalous dimension coefficients $\gamma^q_n$ and $\gamma^g_n$ for quark and gluon-induced processes, the cusp anomalous dimensions, as well as the  $\beta$-function are listed in the appendix of \cite{Becher:2009qa}. The cusp anomalous dimension governs the Sudakov integral
\begin{equation}\label{RGEsols}
\begin{aligned}
   S(\mu_h,\mu) 
   &= \frac{\Gamma_0}{4\beta_0^2}\,\Bigg\{
    \frac{4\pi}{\alpha_s(\mu_h)} \left( 1 - \frac{1}{r} - \ln r \right)
    + \left( \frac{\Gamma_1}{\Gamma_0} - \frac{\beta_1}{\beta_0}
    \right) (1-r+\ln r) + \frac{\beta_1}{2\beta_0} \ln^2 r \\
   &\hspace{1.6cm}\mbox{}+ \frac{\alpha_s(\mu_h)}{4\pi} \Bigg[ 
    \left( \frac{\Gamma_1\beta_1}{\Gamma_0\beta_0} 
    - \frac{\beta_2}{\beta_0} \right) (1-r+r\ln r)
    + \left( \frac{\beta_1^2}{\beta_0^2} 
    - \frac{\beta_2}{\beta_0} \right) (1-r)\ln r \\
   &\hspace{3.7cm}
    \mbox{}- \left( \frac{\beta_1^2}{\beta_0^2} 
    - \frac{\beta_2}{\beta_0}
    - \frac{\Gamma_1\beta_1}{\Gamma_0\beta_0} 
    + \frac{\Gamma_2}{\Gamma_0} \right) \frac{(1-r)^2}{2} \Bigg] + \dots \Bigg\} \,.
\end{aligned}
\end{equation}

\subsection{Anomaly exponent}
\label{app:anom}

In \eqref{combi} in the main text we have combined the anomaly exponent with the double logarithmic part of the beam functions into the exponent  \cite{Becher:2011xn}
\begin{equation}\label{logE}
\begin{aligned} 
   g_i(\eta_i,L_\perp,a_s) 
   &= - \eta_i L_\perp -  a_s \left( \Gamma_0^i + \eta_i\beta_0 \right) \frac{L_\perp^2}{2}-  a_s \left( 2\gamma_0^i + \eta_i\frac{\Gamma_1^i}{\Gamma_0^i} \right) L_\perp
    - a_s^2 \left( \Gamma_0^i+ \eta_i\beta_0 \right) \beta_0\,\frac{L_\perp^3}{3}  \\
   &- a_s \eta_i d_2
    - a_s^2 \left( \Gamma_1^i + 2\gamma_0^i\beta_0 + \eta_i \left( \beta_1 + 2\beta_0\frac{\Gamma_1^i}{\Gamma_0^i} \right) 
    \right) \frac{L_\perp^2}{2} -  a_s^3 \left( \Gamma_0^i + \eta_i\beta_0 \right) \beta_0^2\,\frac{L_\perp^4}{4} \,.
\end{aligned}
\end{equation}
In addition to the various anomalous dimensions and the quantity $\eta_i$ defined in \eqref{etadef}, the exponent involves the anomaly coefficient
\begin{equation}
d_2= \left( \frac{202}{27} - 7\zeta_3 \right) C_A - \frac{56}{27}\,T_F n_f\,.
\end{equation}
As discussed in the main text, the quantity $L_\perp$ has to be counted as $L_\perp\sim \frac{1}{\sqrt{\alpha_s}}$ for very small $q_T$, which explains the presence of higher-order terms multiplying powers of $L_\perp$ in the exponent \eqref{logE}.  These terms ensure that we include terms up to $\mathcal{O}(\alpha_s)$ also in the modified power counting relevant for $q_T \to 0$.

\subsection{Beam functions}
\label{app:b}

We list here the ingredients of the perturbative part $\bar I_{q\leftarrow i}$ in \eqref{barI} of the beam functions for quark induced processes. At one-loop level, these include the standard one-loop Altarelli-Parisi kernels 
\begin{equation}\label{APkernels}
   {\cal P}_{q\leftarrow q}^{(1)}(z) 
   = 4C_F \left( \frac{1+z^2}{1-z} \right)_+ , \qquad
   {\cal P}_{q\leftarrow g}^{(1)}(z) 
   = 4T_F \left[ z^2 + (1-z)^2 \right]\,,
\end{equation}
which multiply the logarithm $L_\perp$ at one loop, as well as the remainder functions
\begin{equation}
   {\cal R}_{q\leftarrow q}(z) 
   = C_F \left[ 2(1-z) - \frac{\pi^2}{6}\,\delta(1-z) \right] , \qquad
   {\cal R}_{q\leftarrow g}(z) 
   = 4T_F\,z(1-z)\,
\end{equation}
obtained in \cite{Becher:2010tm}. To correctly treat the region of very low transverse momentum, we further need the convolutions \eqref{dbarI} of  Altarelli-Parisi kernels, which multiply $L_\perp^2$ at  two-loop order. The results for these quantities are \cite{Becher:2011xn}
\begin{eqnarray}\label{BMfct}
   {\cal D}_{q\leftarrow q\leftarrow q}(z) 
   &=& 16 C_F^2\,\Bigg[ 4 \left( \frac{\ln\frac{(1-z)^2}{z}}{1-z} \right)_+ 
    + 3 \left( \frac{1+z^2}{1-z} \right)_+ - 4(1+z) \ln(1-z) + 3(1+z) \ln z \nonumber\\
   &&\hspace{12mm}\mbox{}- 2(1-z) - \frac94\,\delta(1-z) \Bigg] \,,\\
   {\cal D}_{q\leftarrow g\leftarrow q'}(z) 
   &=& 16 C_F T_F \left[\frac{4}{3z} + 1 - z - \frac{4z^2}{3} - 2(1+z)\ln z
   \right]  ,\nonumber\\
   {\cal D}_{q\leftarrow q\leftarrow g}(z) 
   &=& 16 C_F T_F \left[ \left( z^2 + (1-z)^2 \right) \ln\frac{(1-z)^2}{z} - 2z^2\ln z
    - \frac12 + 2z \right] , \nonumber\\
   {\cal D}_{q\leftarrow g\leftarrow g}(z) 
   &=& 32 C_A T_F \left[ \left( z^2 + (1-z)^2 \right) \ln(1-z) + (1+4z) \ln z
    + \frac{2}{3z} + \frac12 + 4z - \frac{31z^2}{6} \right] \nonumber\\
   &&\mbox{}+ 8\beta_0 T_F \left[ z^2 + (1-z)^2 \right] . \nonumber
\end{eqnarray}

\subsection{NLO expansion of the resummed cross section}
\label{app:nlo}

To $\mathcal{O}(\alpha_s)$, the expansion of the resummed cross section for $q_T>0$ for quark-induced processes is given by
\begin{equation}\label{expNLO}
\begin{aligned} 
\frac{d\sigma_{ij}^{\text{NNLL}}}{dq_T^2} \Big|_{\text{ exp. to NLO}}
&= d\sigma^{0}_{ij} \frac{a_s}{q_T^2} \bigg[\left(C_F\, \Gamma_0 \ln\frac{Q^2}{q_T^2}+2\gamma_0^q\right)
B_i^{(0)}(\xi_1,\mu)B_j^{(0)}(\xi_2,\mu) \\
 &\quad+\frac12 \left(B_i^{(0)}(\xi_1,\mu)B_j^{(2)}(\xi_2,\mu)+B_i^{(2)}(\xi_1,\mu)B_j^{(0)}(\xi_2,\mu)\right)\bigg] \,.
\end{aligned}
\end{equation}
The above result arises from the expansion of ${\cal F}_{ij}$ which starts at $\mathcal{O}(\alpha_s)$ for $q_T>0$. For the hard function we can thus use the leading order result ${\cal H}_{ij}=1$ at this accuracy. In our code, we implement the expanded result as a weight factor exactly as we did for the resummed result, see \eqref{eq:reweight}. 

\end{appendix}

\end{document}